\documentclass[jcp,aip,amsmath,amssymb,floatfix,reprint]{revtex4-2}
\usepackage{pifont}
\usepackage[utf8]{inputenc}
\usepackage[T1]{fontenc}
\usepackage{etoolbox}
\usepackage{graphicx} 
\usepackage{bm} 
\usepackage{wrapfig} 
\usepackage[colorlinks,allcolors=black,citecolor=blue,urlcolor=blue]{hyperref} 
\usepackage{booktabs} 
\usepackage{silence}
\usepackage{multirow}
\usepackage[normalem]{ulem}
\makeatletter
\def\@email#1#2{%
 \endgroup
 \patchcmd{\titleblock@produce}
  {\frontmatter@RRAPformat}
  {\frontmatter@RRAPformat{\produce@RRAP{*#1\href{mailto:#2}{#2}}}\frontmatter@RRAPformat}
  {}{}
}%
\makeatother

\begin{document}

\def\mstitle{Bayesian learning for accurate and robust biomolecular force fields}

\title{\mstitle}

\author{Vojtech Kostal}

\author{Brennon L. Shanks}

\author{Pavel Jungwirth*}
\email{pavel.jungwirth@uochb.cas.cz}

\author{Hector Martinez-Seara*}
\email{hseara@gmail.com}

\affiliation{
Institute of Organic Chemistry and Biochemistry of the Czech Academy of Sciences, Flemingovo nám. 2, 166 10 Prague 6, Czech Republic
}

\begin{abstract}

Molecular dynamics is a valuable tool to probe biological processes at the atomistic level --- a resolution often elusive to experiments.
However, the credibility of molecular models is limited by the accuracy of the underlying force field, which is often parametrized relying on ad hoc assumptions.
To address this gap, we present a Bayesian framework for learning physically grounded parameters directly from ab initio molecular dynamics data.
By representing both model parameters and data probabilistically, the framework yields interpretable, statistically rigorous models in which uncertainty and transferability emerge naturally from the learning process.
This approach provides a transparent, data-driven foundation for developing predictive molecular models and enhances confidence in computational descriptions of biophysical systems.
We demonstrate the method using 18 biologically relevant molecular fragments that capture key motifs in proteins, nucleic acids, and lipids, and, as a proof of concept, apply it to calcium binding to troponin -- a central event in cardiac regulation.

\end{abstract}

\date{\today}
\maketitle

\section*{Introduction}

Molecular dynamics (MD) simulations represent a powerful tool for probing complex biological and other soft matter systems at atomistic resolution~\cite{Dror2012-10.1146/annurev-biophys-042910-155245}.
At this resolution scale, empirical fixed-charge force fields remain the most practical choice to access relevant time and length scales~\cite{Dauber-Osguthorpe2019-10.1007/s10822-018-0111-4}.
Their efficiency comes at the cost of simplifying non-bonded interactions to pairwise Coulomb terms and dispersion-repulsion contributions, typically modeled by the Lennard-Jones potential~\cite{Riniker2018-10.1021/acs.jcim.8b00042}.
While these simplified potential forms enable large-scale simulations, they inherently neglect the explicit electronic structure. To exploit the full power of this approach for accurate modeling of condensed-phase behavior therefore requires careful force field parametrization.

Direct parameterization of an entire biological system based solely on experimental data is generally unfeasible, as the number of adjustable parameters far exceeds the number of available observables, leading to an under-determined optimization problem~\cite{Poleto2022-10.1038/s42004-022-00653-z}.
A widely adopted strategy to overcome this limitation is to decompose the system into smaller, chemically meaningful fragments which are parameterized independently and then recombined.
Established biomolecular force fields are typically developed using a combination of experimental measurements and calculated quantum chemical reference data~\cite{VanDerSpoel2021-10.1016/j.sbi.2020.08.006, Nerenberg2018-10.1016/j.sbi.2018.02.002, Chipot2024-10.1021/acs.jpcb.4c06231}.

Among the various parameters of a force field, atomic partial charges play a particularly critical role.
These charges constitute a coarse-grained representation of the underlying electronic structure and are not themselves experimental observables.
Indeed, there exists a many-to-one mapping between possible partial charge distributions and a given measurable property, making their determination inherently non-unique.
Most established force field development protocols start from gas-phase or partially hydrated  quantum chemical calculations combined with population analyses or charge fitting to reproduce the electrostatic potential to generate initial charge assignments~\cite{Riniker2018-10.1021/acs.jcim.8b00042, Bayly1993-10.1021/j100142a004}.
These values can subsequently be refined to improve agreement with condensed-phase properties.
Various strategies have been proposed to account for environmental influences, including effects of electronic polarization \cite{Leontyev2014-10.1063/1.4884276,Jorge2024-10.1063/5.0236899}, such as minimal explicit hydration~\cite{Vanommeslaeghe2010-10.1002/jcc.21367, Zhu2012-10.1002/wcms.74}, continuum solvent models~\cite{Duan2003-10.1002/jcc.10349} or empirical scaling of charges~\cite{Cerutti2013-10.1021/jp311851r, Duboue-Dijon2020-10.1063/5.0017775a}.
However, these \textit{ad hoc} approaches remain limited in their ability to consistently mimic solvation effects.

Current state-of-the-art force field parametrization methods yield as a rule a single “best” parameter set, leaving little room for modifications without risking uncontrolled degradation of performance.
Experience has taught us that some parameters are more sensitive than others, but this understanding has so far relied largely on trial and error, without systematic evaluation of the ranges within which parameters can be adjusted without compromising accuracy.
Certain flexibility in parameter values can be valuable, for instance when introducing a non-standard residue that must be integrated with an existing structure, potentially creating a net charge that requires redistribution, or when fine-tuning of parameters is required to reproduce an experimental observable such as a ligand binding affinity.
However, making such adjustments reliably is challenging, as they may inadvertently affect important properties of the system—such as hydration structures or hydrogen-bond networks—that are critical for biomolecular stability.
In this light, a method that provides not only the optimal parameter set but also confidence intervals within which the force field can be safely modified would be highly desirable and beneficial to the community.

To depart from the traditional paradigm, we designed a Bayesian framework for force field parameterization that provides rigorous predictions of both optimal parameters and confidence intervals. 
We anchor the force field parameterization to \textit{ab initio} molecular dynamics (AIMD) in explicit solvent, which allows us to naturally capture the effect of the environment without any ad hoc corrections for the gas-to-liquid transition.
Moreover, we leave maximum flexibility by not binding the method to any specific model for water and ions, making it fully general and ready to be applied to any underlying model of choice.

\section*{Results}

\subsection*{Overview of the Method}

Here, we present a general and robust workflow for optimizing partial charge distributions in molecules within the framework of fixed-charge models.
Unlike for neat solvents or simple ion solutions, experimental data directly applicable for parametrization are rarely available for biomolecular species, making it difficult to anchor parameter optimization to the closest ground truth data and its associated deviations.
To circumvent this limitation, we focus on condensed-phase reference data obtained from AIMD simulations based on density functional theory, which have become more accessible due to improvements in computational methods and infrastructure.
These simulations inherently capture many-body interactions, including electronic polarization, thus providing a physically realistic structural benchmark at an acceptable computational cost.
By linking parameter optimization to high-level simulations of the target environment, we ensure that the resulting force field parameters accurately reproduce the condensed-phase behavior of the molecules within the accuracy limits imposed by the force field's functional form and its quantum mechanical reference.

\begin{figure*}[ht!]
    \centering
    \includegraphics[width=0.99\linewidth]{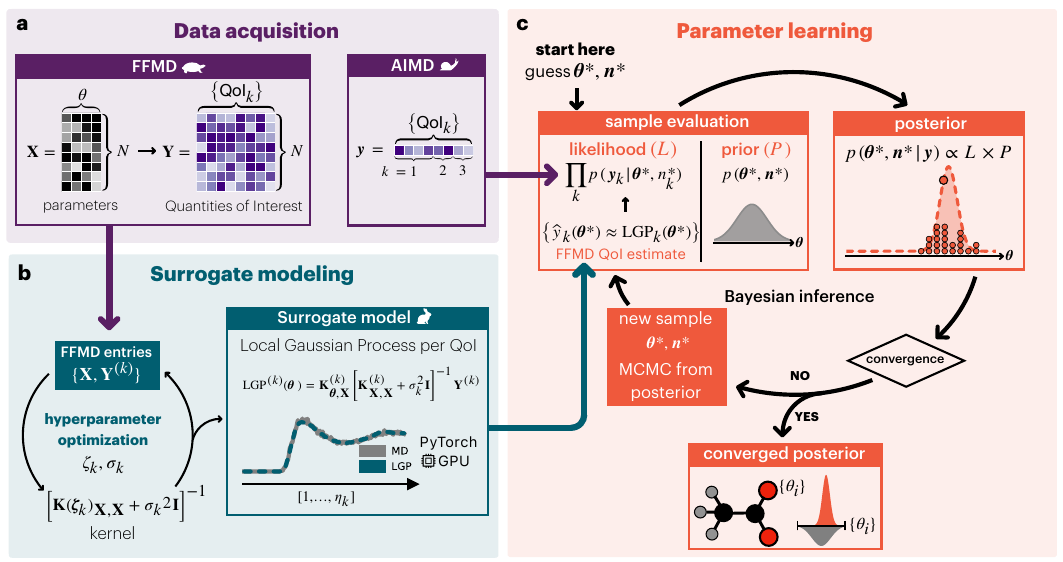}
    \caption{
    \textbf{Overview of the Bayesian inference workflow.}
    The workflow begins with data acquisition (a): a set of $N$ randomized partial charge vectors $\boldsymbol{\theta}$ collected in matrix $\mathbf{X}$ are used to run FFMD simulations and the corresponding QoI outputs are stored in matrix $\mathbf{Y}$.
    AIMD simulations provide reference trajectories and QoIs are extracted into vector $\boldsymbol{y}$.
    The training output matrix $\mathbf{Y}$ is partitioned column-wise into sub-matrices $\mathbf{Y}^{(k)}$ corresponding to the $k$-th QoI.
    In surrogate modeling (b), a separate LGP is trained for each QoI.
    Kernel hyperparameters are optimized to the FFMD dataset $\{\mathbf{X}, \mathbf{Y^{(k)}}\}$ using leave-one-out marginal likelihood.
    The resulting inverse kernel matrix is precomputed to construct the final LGP surrogate.
    In parameter optimization (c), Bayes' theorem combines a prior probability with the likelihood of observing the AIMD reference features $\boldsymbol{y}$ given predicted FFMD QoIs from the LGPs.
    Posterior sampling is performed using Markov Chain Monte Carlo (MCMC).
    In each iteration, new parameters $\boldsymbol{\theta}^*$ and QoI-specific nuisance parameters $\boldsymbol{n}^*$ are proposed, FFMD QoIs $\boldsymbol{y}(\boldsymbol{\theta}^*)$ are predicted from the set of surrogates, and the likelihood is evaluated.
    Multiplying the likelihood by the prior yields the posterior value up to a normalization constant.
    The MCMC loop continues until the posterior distribution converges, yielding the optimized force field parameter distribution.
    }
    \label{fig:optimization-chart}
\end{figure*}

The core of our approach is a generalizable method for learning partial charge distributions of molecular fragments, readily extensible to other force field parameters through an automated data acquisition and accelerated Bayesian inference framework (Fig.~\ref{fig:optimization-chart}).
Our method learns classical force field parameters from AIMD simulations of solvated molecular fragments, thereby incorporating solvent-mediated polarization effects that are pervasive in condensed phase systems. 
In this approach, force field molecular dynamics (FFMD) simulations with trial parameters were used to generate quantities of interest (QoIs), such as radial distribution functions and hydrogen bond order (Fig.~\ref{fig:optimization-chart}1), which were then emulated by local Gaussian process (LGP) surrogate models (Fig.~\ref{fig:optimization-chart}2). 
The LGP surrogate predictions enabled efficient evaluation of the likelihood of candidate parameter sets against reference AIMD data, and when combined with prior information, the approach was used to learn the posterior distribution with Markov chain Monte Carlo, as schematically depicted in Figure~\ref{fig:optimization-chart}. 
By iteratively sampling from this posterior, the method refined the parameter estimates until convergence, yielding charge distributions that balance accuracy and uncertainty inherent to the Bayesian formulation.

We emphasize that the computational tractability of our method hinges on the LGP regressor~\cite{Shanks2024-10.1021/acs.jctc.3c01358}.
This surrogate predicts the structural QoIs from a given set of trial charges at a fraction of the time of an FFMD simulation, allowing for efficient evaluation of new candidates during Markov chain Monte Carlo sampling without the need to repeatedly run costly FFMD simulations. 
LGPs were favored over black-box machine learning tools, such as neural networks, since they can incorporate interpretable, physics-informed priors that effectively model structural properties in bulk liquids~\cite{Shanks2024-10.1021/acs.jpclett.4c02941} and aqueous ions~\cite{Fan2025-10.1021/acs.jctc.5c00873}.

We demonstrate the strength and generality of our methodology by optimizing charge distributions for a set of biologically relevant molecular species ranging from neutral, cationic, to anionic motifs, including carboxylates, phosphates, sulfates, carbonyls, and positively charged nitrogen centers characteristic of protein sidechains, lipid headgroups, nucleic acids, and glycans.
As a proof-of-principle, we then apply the acetate parameterization to model calcium binding to troponin --- a key event regulating the heartbeat that is challenging to simulate due to the large number of charged groups and the calcium ion involved. 
Together, these results establish a systematic strategy for deriving fragment-based force field parameterizations that bridge electronic-structure accuracy and classical simulation efficiency, with broad applications in biomolecular modeling.

\subsection*{Parameterization of Partial Charge Distributions}

\begin{figure*}
    \centering
    \includegraphics[width=\linewidth]{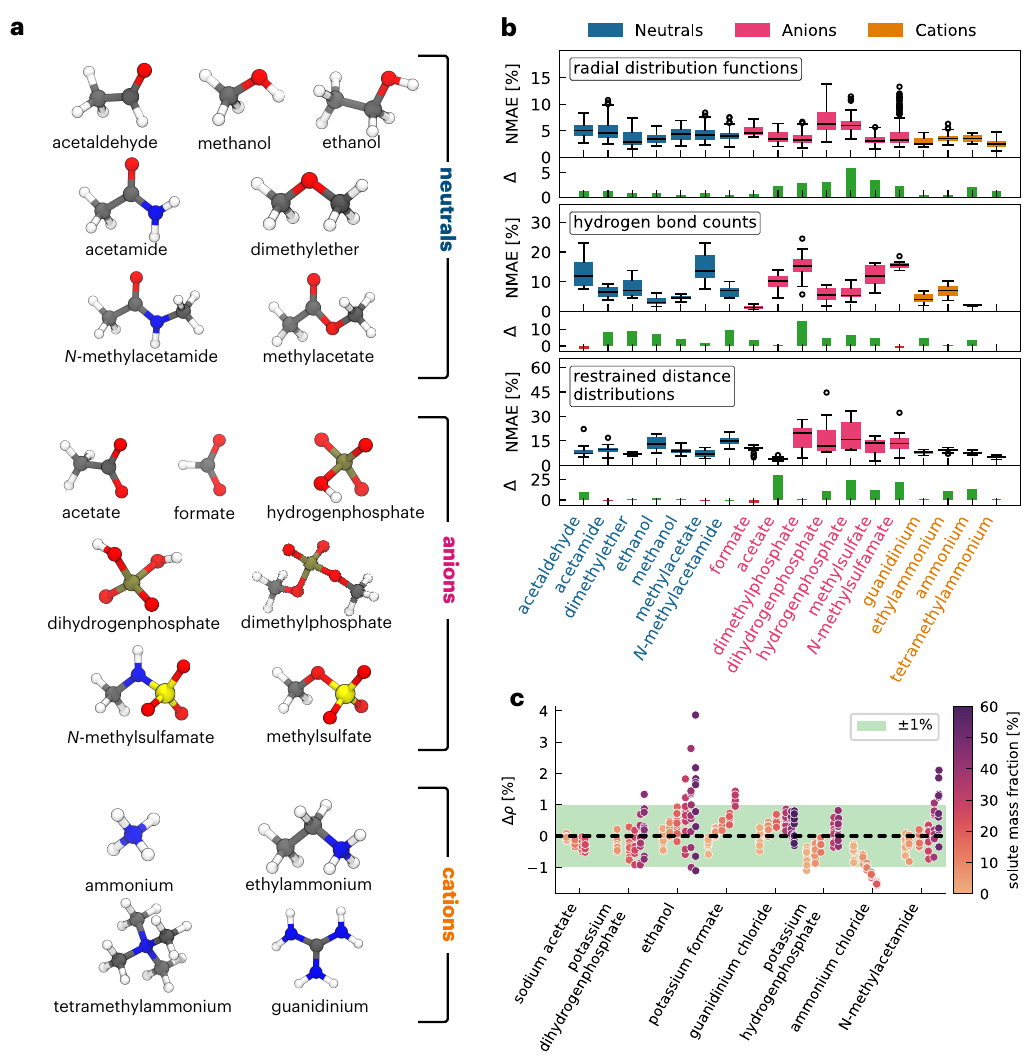}
    \caption{
        \textbf{Accuracy and validation of optimized partial charges.} 
        \textbf{a}: Licorice representations of the parameterized species grouped according to their net charge.
        \textbf{b}: Boxplots showing NMAE of ten samples from the partial charge posterior distribution against AIMD references for three QoIs, color-coded by molecular charge: neutral (blue), anions (pink), and cations (orange).
        Subpanels indicate relative average improvement in percentage with respect to CHARMM36-nbfix: green - improvement, red - regression.
        \textbf{c}: Relative deviations ($\Delta$) between simulated and experimental densities for selected species at different concentrations 298~K.
        The green band indicates the $\pm1\%$ error margin.
    }
    \label{fig:accuracy}
\end{figure*}

We applied the Bayesian inference framework to refine partial charge distributions for 18 biologically important motifs, illustrated in Figure~\ref{fig:accuracy}a.
Starting from the CHARMM36 baseline, we learned the partial charges while leaving bonded and van der Waals parameters unchanged -- a decision that affects the resulting optimized charges.
Here, we applied the electronic continuum correction (ECC) as a baseline framework for charges that includes electronic polarization in a mean field manner~\cite{Leontyev2010-10.1021/ct9005807b, Leontyev2011-10.1039/C0CP01971Ba, Kirby2019-10.1021/acs.jpclett.9b02652a, Duboue-Dijon2020-10.1063/5.0017775a, Kostal2023-10.1021/acs.jpclett.3c02231}.
We enforced a global constraint on the total molecular charge, scaled by a factor of 0.8.
This scaling factor was adopted from our recent work, together with compatible models of water~\cite{CrucesChamorro2024-10.1021/acs.jpclett.4c00344} and ions~\cite{Fan2025-10.1021/acs.jctc.5c00873}.

For each molecule, the training set comprised several thousand charge distributions sampled from a physically motivated truncated normal prior.
Charge bounds were chosen based on ranges typical of established force fields and extended to allow broader, yet physically meaningful, variation.
QoIs were extracted for each charge distribution from three independent simulation setups: (i) the aqueous solute, (ii) the solute with a counterion in direct contact, and (iii) the solute with a solvent-shared counterion.
After training an LGP surrogate to map partial charges to the QoIs, the Markov chain Monte Carlo engine was used to sample the posterior distribution of partial charges.
The accuracy of the QoIs relative to the AIMD reference was assessed using the normalized mean absolute error (NMAE) from ten posterior samples with different partial charge distributions.

Figure~\ref{fig:accuracy}b summarizes the overall performance of the method.
The optimized charges show consistent agreement with AIMD across all metrics.
Hydration structure errors, characterized by radial distribution functions (RDFs), remain below 5\% for most species, demonstrating robust reproduction of solvation structure.
Hydrogen bond counts typically deviate by less than 10–20\%, with larger variability arising from the rigid water model, which to a certain extent limits simultaneous reproduction of RDFs and hydrogen-bond statistics.
Ion-pair distance distributions generally show deviations under 20\%, with somewhat larger errors for some anionic species.
The residual errors largely reflect the trade-off inherent in simultaneously reproducing all the QoIs.
For comparison, Supplementary Figure~S3 shows the performance obtained with parameters optimized individually for each QoI.

Relative improvements compared to the original CHARMM36-nbfix force field are quantified in the subpanels of Figure~\ref{fig:accuracy}b by green and red bars.
Systematic improvements are observed across nearly all species and QoIs, particularly for charged systems --- most notably anions --- where optimized charges restore more balanced electrostatics.
Even neutral molecules benefit modestly.
While only the mean trends are shown for clarity, individual posterior samples often achieve substantially greater improvements than reflected in the averages.
A full distribution of $\Delta$ values, including variability and outliers, is provided in Supplementary Figure~S5.

As the parametrization was performed for system sizes and concentrations accessible to AIMD, direct transferability to higher solute concentrations is not automatically guaranteed because explicit solute-solute interactions are missing in the training and reference data.
To assess this potential effect, we validated our charge distributions against known aqueous solution densities using posterior charge samples, as shown in Figure~\ref{fig:accuracy}c.
For charged species, counterions (sodium, potassium or chloride) were selected based on available experimental data, and the corresponding concentration ranges extended up to their experimental solubilities.
Across nearly all solutions, simulated densities deviated from experiment by less than 1\%, even at solute concentrations approaching a 50\% mass ratio.
This agreement --- despite no explicit training on solute–solute interactions --- provides strong evidence for the robustness and transferability of the optimized charges.

\subsection*{Chemical Intuition - Learning from the Posterior}

\begin{figure*}
    \centering
    \includegraphics[width=\linewidth]{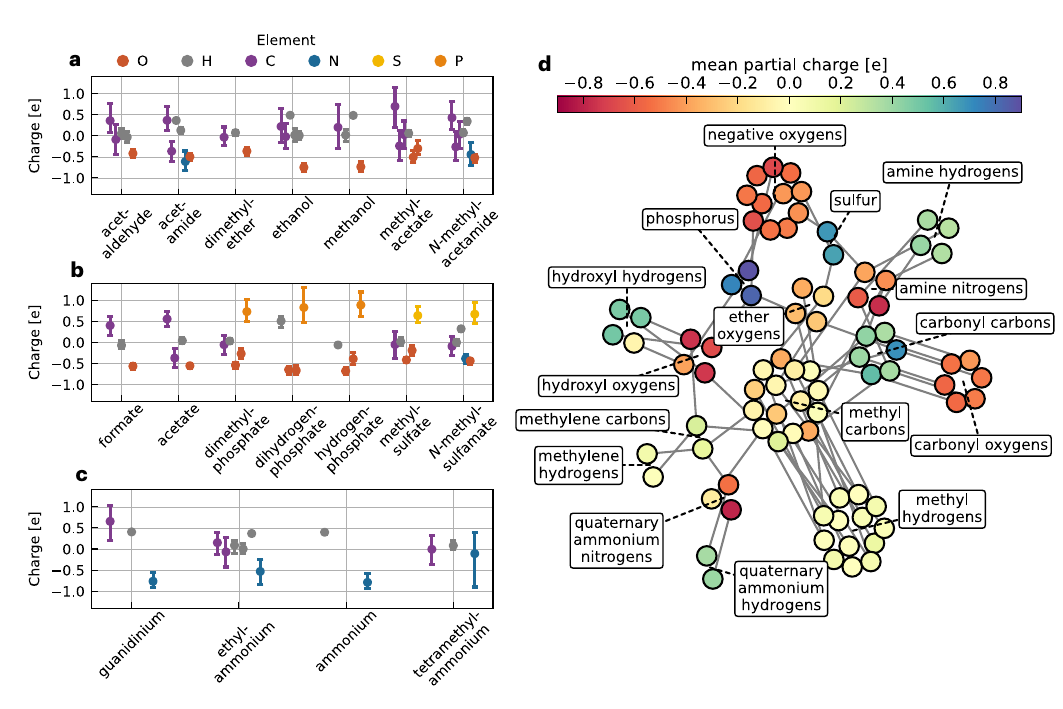}
    \caption{
    \textbf{Chemical insights from the Baysian inference of partial charge distributions.}
    \textbf{a-c}: posterior mean partial charges (bullets) with 95\% confidence intervals (errorbars) for all atom types across the set of species parameterized in this work, color-coded by chemical element.
    The species are grouped by their net charge into individual panels as (a) neutrals, (b) anions, and (c) cations.
    \textbf{b}: Graph representation of the atomic partial charges grouped based on their chemical similarity.
    Nodes of the graph represent the individual atomtypes, while the edges present the covalent bonds if there is such a physical connection.
    }
    \label{fig:optimized-charges}
\end{figure*}

Building on the validation of optimized charges against AIMD and experimental data, we next examine the chemical insight provided by the Bayesian framework.
Figures~\ref{fig:optimized-charges}a-c present the posterior means and associated 95\% confidence intervals for all atoms across the studied molecules.
These intervals, known as 1D marginals of the partial charge joint probability distributions, naturally serve as a quantified form of chemical intuition by delineating ranges of charge values that preserve consistency with the reference AIMD data.
In this way, the posterior charge distributions serve as a robust design space for further tuning and sensitivity analysis.

For example, atomic partial charges with narrow confidence intervals such as is the case for oxygen atoms (red-colored in Figure~\ref{fig:optimized-charges}a) are tightly defined by the structural QoI. 
Thus, they play a critical role in reliably representing the reference data. 
In contrast, broader confidence intervals (\textit{e.g.}, for carbon or phosphorus atoms) indicate that these atoms have a less critical influence on the QoIs and can accommodate more variability without degrading agreement with the reference.
This qualitative observation agrees with chemical intuition: atoms directly exposed to the environment are more important in reproducing solvation structure than those buried deeper within the molecule and shielded by other atoms.

Figure~\ref{fig:optimized-charges}d offers a graph-based perspective of the optimized charges. 
The graph layout was generated using community detection and force-directed placement, providing both topological and chemical organization.
Atoms are clustered based on chemical similarity into functional groups, with nodes representing atomtypes (element plus bonding environment) and edges indicating covalent connectivity.
Node colors encode the most probable partial charge, spanning approximately $-0.8e$ to $+0.9e$, a narrower range than in the CHARMM36-nbfix force field ($-1.0e$ to $+1.5e$), in part due to the ECC scaling factor.

Despite optimizing charges independently for each molecule, clear chemical trends emerge.
Strongly negative atom types correspond to oxygen or nitrogen atoms.
Terminal oxygens in carboxylates and phosphates cluster between $-0.65e$ and $-0.55e$, while sulfates are slightly less negative ($\sim -0.4e$), reflecting charge delocalization over three equivalent atoms.
Hydrogenphosphate oxygens reach about $-0.67e$.
Carbonyl oxygens span $-0.7e$ to $-0.4e$, consistent with their terminal, non-resonance-stabilized positions, and hydroxyl oxygens are typically around $-0.7e$, except in hydrogenphosphate ($-0.4e$), likely due to nearby electronegative sites.
Ether oxygens (bound to two heavy atoms) are less polarized ($-0.3e$ to $-0.1e$).
Nitrogens display a broad spread ($-0.9e$ to $-0.4e$), except for quaternary nitrogen in tetramethylammonium, which can also hold positive values, reflecting strong electron withdrawing effects of methyl groups.
Hydrogens bound to electronegative atoms (O or N) are significantly positive, while aliphatic hydrogens are weakly positive, with a few exceptions showing small negative values.
Carbon atomtypes show the widest charge variation.
Methyl carbons are weakly negative but can reach up to $-0.4e$, methylene carbons are typically positive ($+0.2e$), and carbonyl carbons are strongly positive, consistent with electron-withdrawing effects from the resonant oxygens.
These trends closely align with chemical intuition, supporting the physical plausibility of the optimized charges.

Consistency across independent molecule-specific optimizations demonstrates the robustness of the Bayesian framework, which not only fits reference data but also captures generalizable chemical principles with rigorous uncertainty quantification. 
This makes the approach highly promising for transferable parameterization of molecular fragments in condensed-phase simulations, as demonstrated in the next section. 
Indeed, knowledge of the plausible confidence intervals can be used to incorporate small fragments into larger molecules or more complex environments, for instance, by indicating which atoms can absorb any excess of charge while minimizing perturbations to the QoIs. 
Finally, parameters with higher uncertainty can be identified and more aggressively refined by selecting new or additional reference data. 
This strategy not only guides future parameterization efforts but also provides critical insight into how the force field can be further improved.

\begin{figure*}[t!]
    \centering
    \includegraphics[width=\linewidth]{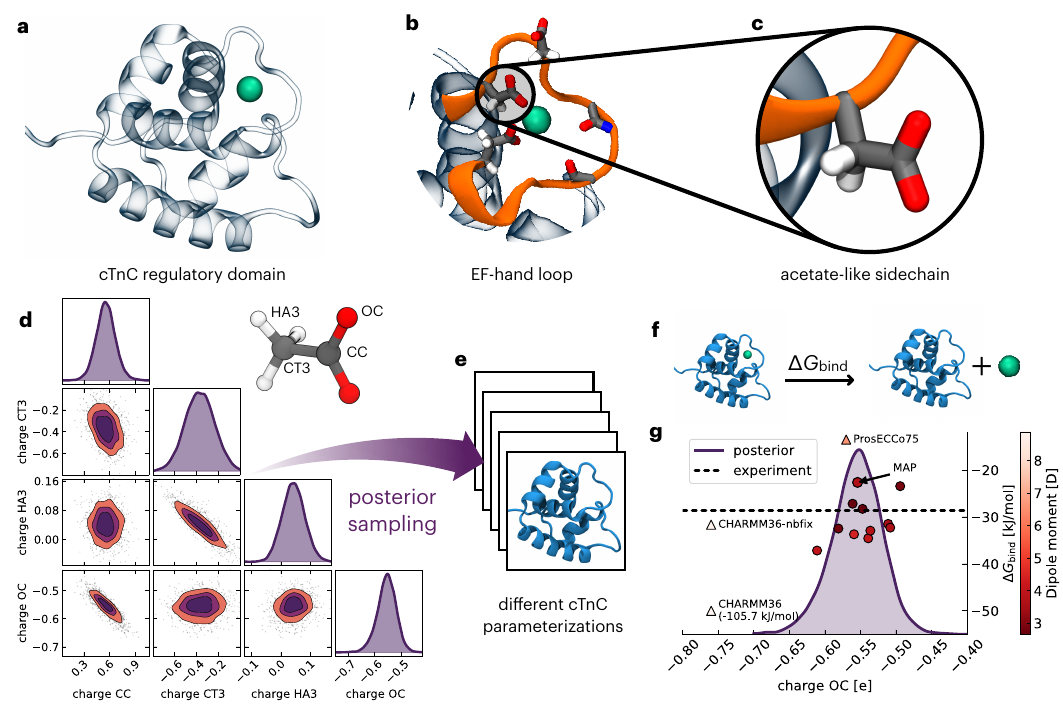}
    \caption{
    \textbf{Calcium binding to the regulatory domain of human cardiac troponin-C (N-cTnC).}
    \textbf{a-c}: Cartoon representations of the relevant parts of the N-cTnC.
    \textbf{a}: Overall structure of N-cTnC (transparent blue) with bound Ca$^{2+}$ (green).
    \textbf{b}: Close-up of the EF-hand loop (orange) highlighting Ca$^{2+}$-coordinating residues (licorice representation)
    \textbf{c}: Zoom-in of the characteristic carboxylate motif involved in Ca$^{2+}$ binding.
    \textbf{d}: Posterior distributions of the optimized partial charges of acetate shown on the diagonal (purple), with pairwise parameter correlations on the off-diagonals (density contours).
    Individual posterior samples are indicated as gray background dots.
    \textbf{e}: Protein parameterizations generated by sampling carboxylate charges from the acetate posterior distribution.
    \textbf{f}: Schematic depiction of the unbound (left) and Ca$^{2+}$-bound (right) states of N-cTnC used to compute the binding free energy.
    \textbf{g}: Computed Ca$^{2+}$ binding free energies for N-cTnC as a function of the sampled carboxyl oxygen charge with its marginal posterior (purple).
    Points are colored by the dipole moment of the CH$_2$COO$^-$ fragment.
    Bullets show results from the force fields developed in this work, while triangles represent values from CHARMM36, CHARMM36-nbfix, and ProsECCo75.
    The experimental $\Delta G_\text{bind}$ (–28.6~kJ/mol) is indicated by the dashed black line.
    }
    \label{fig:troponin}
\end{figure*}

\subsection*{Proof-of-Principle: Calcium Binding to Cardiac Troponin C}

We leverage the fragment-level chemical intuition provided by the Bayesian posterior to demonstrate how our method can be successfully applied in a biologically relevant context.
To this end, we evaluated the binding free energy of Ca$^{2+}$ to the regulatory domain of human cardiac troponin C (N-cTnC) (Figure~\ref{fig:troponin}a), which is the primary Ca$^{2+}$ sensor that regulates cardiac contraction.

To simulate the entire protein, a complete and self-consistent set of force field parameters was required. 
For N-cTnC, this corresponds to an inherently high-dimensional parameter learning problem, which we addressed using a fragment-based Bayesian approach.
Keeping the protein backbone parameters fixed, all charged side chains --- carboxymethyl, guanidinium, and ethylammonium --- were parameterized within this framework to ensure internal consistency, thereby reducing the dimensionality of the parameter space.

Partial charges of the carboxymethyl side chains, the principle charged motif of the Asp and Glu residues in the EF-hand loop of N-cTnC (Figure~\ref{fig:troponin}a–c), were modeled using samples drawn from the posterior distribution of acetate (Figure~\ref{fig:troponin}d–e).
The positively charged side chain of arginine was parameterized using guanidinium, while the lysine side chain was derived from ethylammonium.
For a schematic breakdown of the amino acid side chains, see Supplementary Figure~S6.
A uniform charge scaling factor of 0.8, consistent with the ECC framework~\cite{Fan2025-10.1021/acs.jctc.5c00873}, was applied to all charged residues and incorporated directly during parameterization.
To maintain overall protein electroneutrality, a small compensating charge (up to $\pm0.1e$) was added to or subtracted from the atom with the largest posterior marginal uncertainty, typically a buried carbon atom. 

Distinct sets of partial charges, including the \textit{maximum a posteriori} (MAP), were sampled from each fragment’s posterior distribution, yielding multiple N-cTnC force field parametrizations (Figure~\ref{fig:troponin}e and Supplementary Table~S1).
Binding free energies of Ca$^{2+}$ to N-cTnC obtained with the MAP and ten different partial charge sets are shown in Figure~\ref{fig:troponin}d as bullets.
We see a correlation between $\Delta$G$_\text{bind}$ and partial charge of the carboxyl oxygen, which is the leading direct binding partner of Ca$^{2+}$ in the EF-hand loop.
The predicted values are in good agreement with the experimental value of $-28.6$~kJ/mol~\cite{Rayani2021-10.1016/j.jbc.2021.100350}, deviating by less than 10~kJ/mol, with certain parameter sets achieving near-quantitative accuracy of $\sim$1~kJ/mol. 

Some samples, however, exhibit minor over- or underbinding.
Specifically, the MAP parameterization underbinds by $\sim6$~kJ/mol.
This indicates that the partial charge distribution that best represents the acetate ion in bulk water is not necessarily optimal in this biological context.
This observation is not surprising, as the acetate moiety is not identical to the carboxymethyl pattern found in amino acid sidechains.
More broadly, this exemplifies a central limitation of fixed-charge force field optimizers; namely, their inability to adapt seamlessly to new chemical contexts.
This observation thus underscores the importance of the Bayesian approach: by capturing parameter uncertainties, we obtain a more complete and nuanced view of the model’s predictive capabilities.

For comparison, we also evaluated the binding free energies using several established force fields — CHARMM36/CHARMM36-nbfix~\cite{Huang2017-10.1038/nmeth.4067}, and ProsECCo75~\cite{Nencini2024-10.1021/acs.jctc.4c00743} — shown as triangles in Figure~X.
CHARMM36 yields a substantially overestimated binding free energy of $-105.7$~kJ mol$^{-1}$, primarily due to its neglect of electronic polarization.
ProsECCo75, which is a CHARMM36 derivative that approximates polarization effects in a mean-field manner through the ECC scheme, instead shows pronounced underbinding of $-13.5$~kJ/mol.
The CHARMM36-nbfix variant yields a value of $-31.7$~kJ/mol which is close to the experiment.
However, this agreement arises from compensation of errors, namely from empirical, pair-specific adjustments to van der Waals radii to counteract overly strong ion–ion electrostatic interactions.
Interestingly, we observe here a correlation between accuracy and dipole moment of the CH$_2$COO$^-$ motif: sets with dipole moments of 2.7-4.4~D around the most probable oxygen charge region ($\sim$0.57) best reproduce the experimental binding free energy, as illustrated in Figure~\ref{fig:troponin}g.
These dipole moments are markedly smaller than CHARMM36/CHARMM36-nbfix (8.9~D) and prosECCo75 (6.6~D).

\section{Discussion}

Empirical force field design for biomolecules is often as much art as science. The high-dimensional parameter landscapes, sensitivity to loss-function definitions, and dependence on available reference data can lead to markedly different models, often without clear justification or enough information to declare a clear winner. This lack of transparency has been increasingly recognized in the biomolecular modeling community~\cite{Amaro2025-10.1038/s41592-025-02635-0} and has prompted calls for rigorous assessments of both uncertainty and bias~\cite{VanDerSpoel2021-10.1016/j.sbi.2020.08.006}.

Bayesian inference offers a principled route to address these concerns, providing a gold-standard framework for quantifying uncertainty given limited information~\cite{Gelman1995-, VanDeSchoot2021-10.1038/s43586-020-00001-2}.
Its primary drawback is computational: the curse of dimensionality and the cost of repeated evaluations often render Bayesian methods impractical for complex, high-dimensional tasks. 
Consequently, the adoption of Bayesian inference in force field development has been limited, with prohibitive computational costs often outweighing the clear benefits of rigor and interpretability.

In the spirit of transparency and reproducibility, we combined machine learning accelerated Bayesian inference with GPU computing to create an efficient, transferable, and physically grounded tool for learning fixed-charge force field parameters directly from condensed-phase AIMD reference data. 
This approach enables parameter inference at scales and system complexities beyond what was previously feasible. 
As an example presented here, we extended Bayesian learning of classical force fields from bulk liquids-argon ~\cite{Angelikopoulos2012-10.1063/1.4757266}, neon ~\cite{Shanks2024-10.1021/acs.jpclett.4c02941}, and water~\cite{Dutta2018-10.1063/1.5030950}-to 18 biologically relevant molecular fragments in water, increasing parameter dimensionality from 3–5 in prior studies up to 10. 
While the focus of this paper is on the optimization of charges in an explicit-solvation environment, the same methodology is also directly applicable to other force field parameters. 
As a matter of fact, exploratory work optimizing Lennard-Jones non-bonded parameters within the same framework points to additional improvements in accuracy in aqueous solutions.

An essential aspect of our parametrization strategy is its transferability from small molecular fragments in simple environments to more complex and biologically relevant systems. 
We first demonstrated this by reproducing the experimental densities of aqueous electrolyte solutions with charges derived from minimal fragments, achieving errors below 1\% in most cases. 
Next, we extended the same parametrization to the charged residues of the regulatory domain of cardiac troponin C and calculated the free energy of Ca$^{2+}$ binding to its EF-hand loop. 
The close agreement with experimental binding data provides strong evidence that parameters optimized on isolated fragments remain predictive when applied to a protein environment that differs substantially from the training set. 
Moreover, by sampling directly from the posterior distribution, we were able to assess the robustness of alternative parameter sets without further refinements. 
Our approach does not yield a single fixed solution but rather a family of statistically consistent models, leaving users the flexibility to incorporate additional criteria into their choice of parameters. 
Together, these results underscore the robustness and transferability of the approach across both condensed-phase solution properties and ion–protein interactions.

The generality of our framework makes it applicable to a wide range of molecular modeling challenges. 
Because any observable accessible from simulation can be incorporated into the likelihood, the same methodology can unify diverse data types, from \textit{ab initio} calculations to experimental measurements such as thermodynamic properties and neutron/X-ray scattering patterns.
Our method therefore opens the possibility of systematically integrating experimental data with simulation-based models while preserving rigorous uncertainty quantification. 
In this sense, the approach provides a bridge between physics-based modeling and modern data-driven methodologies, offering a common Bayesian language for reconciling calculations and experiment. 
Importantly, the Bayesian framework limits bias from hand-tuned parameters and makes the impact of modeling choices explicit, transparent, and measurable. 
Our method thus offers a principled framework for rigorously evaluating modeling assumptions in relation to target observables.

While our approach substantially broadens the applicability of Bayesian inference to force fields, it is not without limitations. 
The curse of dimensionality poses challenges for scaling to very large biomolecular systems, which may involve hundreds or thousands of parameters, even when leveraging GPU acceleration and machine learning surrogate models.
As a result, practical applications to large-scale force field reparameterization will likely require hybrid strategies that integrate our fragment-based training with targeted refinements for system-specific properties. 
A further limitation is that the posterior distribution remains conditional on the modeling choices used to construct the likelihood, priors, and surrogate models.
If these ingredients are mis-specified, for example due to approximations in surrogate accuracy or oversimplified physical assumptions, the resulting uncertainty estimates may underestimate or overestimate the true parametric uncertainty. 
In addition, the quality of inference is inherently tied to the availability and representativeness of the reference data. 
Sparse, noisy, or biased observables can lead to skewed posterior distributions, particularly as the dimensionality of the parameter space increases.

Despite these limitations, the significant advances demonstrated with our method suggest a new paradigm for force field development: one in which parameter optimization is no longer an art, but rather a transparent and quantitative science. 
By leveraging GPU acceleration and surrogate modeling, we make parameter inference not only rigorous but also practical at scales relevant to biomolecular research. 
The release of our methodology as an open and extensible software package democratizes access to uncertainty-aware parameterization, lowering the barrier for broad adoption and addressing the communities calls for transparency and reproducibility. 
We envision that this framework will help establish uncertainty quantification as a standard practice in molecular simulations, fundamentally reshaping how the community develops, validates, and applies biomolecular force fields.

\section{Methods}

This section outlines the computational framework developed for refining partial atomic charges in classical force fields using structural observables obtained from AIMD simulations.
We begin by introducing the key principles of Bayesian inference that underpin the inference process (Fig \ref{fig:optimization-chart}).
The subsequent sections describe the full optimization workflow, including a detailed description of the surrogate modeling approach based on LGPs, which enables efficient approximation of classical MD outcomes.
Eventually, we provide details of both the DFT- and FFMD simulations used throughout this study.

\subsection{Inference Workflow}
\label{sec:optimization_workflow}

The workflow for optimization of partial charges comprises of four steps depicted in Figure~\ref{fig:optimization-chart}:
(a) data acquisition, (b) surrogate model training, (c) Bayesian inference, and (d) optional validation (not shown in the figure).
The goal is to align structural QoI from FFMD with reference AIMD.
While extension to van der Waals parameters is straightforward and implemented in the code, this study focuses on charge refinement.

\paragraph{\textbf{Data acquisition}}
\label{sec: data_acquisition}

Initial systems (\textit{i.e.}, molecules in aqueous solution) are prepared using user-supplied topologies and equilibrated with FFMD.
The equilibrated snapshots seed AIMD simulations.
In parallel, a diverse FFMD training dataset is generated by sampling the force field parameters $\boldsymbol{\theta}$ via Latin Hypercube Sampling (LHS) implemented in \texttt{SciPy}~\cite{Virtanen2020-10.1038/s41592-019-0686-2}.
The parameter samples are stored in matrix $\mathbf{X}=\left[ \boldsymbol{\theta}_1, \boldsymbol{\theta}_2, \dots, \boldsymbol{\theta}_N \right]^\top \in \mathbb{R}^{N\times D}$, where $N$ is the number of samples and $D=\dim(\boldsymbol{\theta})$.
The sampling is limited to chemically sound regions of the parameter space.
To maintain the total molecular charge, one atom type per molecule is designated as implicit, with its charge calculated by subtracting the sum of all sampled charges from the target total molecular charge.

QoIs are computed from three simulations at both FFMD and AIMD level: solute in water, solute with a restrained nearby ion, and solute with a restrained distant ion.
Structural descriptors (QoI) computed include:
\begin{enumerate}
    \item \textbf{Partial radial distribution functions (RDF)s:} Between solute atoms and water oxygens.
    \item \textbf{Hydrogen bond statistics:} Based on O/N/S acceptor atoms using geometric criteria of the donor-acceptor distance $<3.5$~\AA\ and donor–hydrogen–acceptor angle $>150^{\circ}$.
    \item \textbf{Ion--solute distance probability density:} For restrained ion placements for trajectories containing extra ion.
\end{enumerate}
The FFMD QoIs are then stored in a block-concatenated matrix $\mathbf{Y}=\left[\mathbf{Y}^{(1)} \mathbf{Y}^{(2)} \dots \mathbf{Y}^{(M)} \right]$, where $\mathbf{Y}^{(k)} \in \mathbb{R}^{N\times \eta_k}$ contains $k$-th QoI evaluated at $\eta_k$ grid points.
Similarly, the AIMD QoIs are stored in the vector $\boldsymbol{y}$.
FFMD matrices $\mathbf{X}$ and $\mathbf{Y}$ are used as inputs and outputs in the subsequent surrogate modeling.

\paragraph{\textbf{Surrogate modeling}}

A central challenge in our Bayesian inference framework is the need to evaluate the likelihood function for each parameter set proposed by the Markov chain Monte Carlo sampler.
This evaluation, in principle, requires generating and analyzing a full FFMD trajectory for every candidate set of partial charges $\boldsymbol{\theta}$-a computationally prohibitive task given the tens to hundreds of thousands of such evaluations required for posterior convergence.

\textbf{Local Gaussian Processes:} To overcome this bottleneck, we introduce a surrogate model that maps $\boldsymbol{\theta}$ directly to structural QoIs derived from FFMD simulations.
This surrogate model replaces the expensive FFMD simulations during inference while retaining predictive accuracy.
We construct this surrogate using an LGP regressor~\cite{Shanks2024-10.1021/acs.jctc.3c01358}, which is an efficient approximation to a Gaussian process (GP) that decomposes high-dimensional input space into a subset of independent one-dimensional GPs.
Each QoI (\textit{e.g.}, RDFs or scalar descriptors) is predicted component-wise, with each dimension modeled by a separate GP.
For a structural quantity such as an RDF, this approximation amounts to learning a set of RDF values $\{g(r_i)|i \in 1,\dots,\eta_k\}$ at radial inducing points $r_1,\dots, r_{\eta_k}$, $g(r_i)\sim GP_i$, as opposed to modeling the full RDF as a single GP, $g(\boldsymbol{r}) \sim GP$.

For a trial input $\boldsymbol{\theta}$, the zero mean LGP expectation value of the $k$-th QoI is given by a column vector of the LGP expectation value at each inducing point $i \in 1,\dots, \eta_{k}$:
\begin{equation}
    \label{eq:lgp}
    \widehat{\boldsymbol{y}}_k(\boldsymbol{\theta})
    \approx \text{LGP}_k(\boldsymbol{\theta})
    =
    \mathbf{K}_{\boldsymbol{\theta},\mathbf{X}}^{(k)}
    \left[ \mathbf{K}_{\mathbf{X},\mathbf{X}}^{(k)} + \sigma_k^2\mathbf{I} \right]^{-1}
    \mathbf{Y}^{(k)}
\end{equation}
where $\sigma_k^2$ is the noise variance hyperparameter, and $\mathbf{I}$ is the identity matrix.
This yields a vector prediction $\widehat{\boldsymbol{y}}_k(\boldsymbol{\theta}) \in \mathbb{R}^{\eta_k}$.

To solve Eq.~\eqref{eq:lgp}, we must specify a LGP prior to specify the kernel (or Gram) matrix $\mathbf{K}$ in the form of a kernel function, $K$.
This function defines the underlying properties of samples drawn from the LGP and can be tailored to encode generic function properties such as smoothness, continuity, periodicity, and others~\cite{Rasmussen2008-}.
Here, we choose the squared-exponential (radial basis function) kernel to construct the elements of $\mathbf{K}$ between arbitrary sample indices $i$ and $j$:
\begin{equation}
\label{eq:se-kernel}
K_{ij}^{(k)} = \alpha_k^2 \exp\left[ -\frac{1}{2} \sum_{d=1}^{D} \left( \frac{\theta_{d,i} - \theta_{d,j}}{\ell_{k,d}} \right)^2 \right],
\end{equation}
where $\alpha^2$ is the kernel variance which controls the vertical scale of the prediction, and $l_d$ are length scales that control how quickly the correlation between function values decay with distance. The set of variables  $\boldsymbol{\zeta_k} = \{\sigma_k, \alpha_k, \ell_{k,1}, \dots, \ell_{k,D}\}$ are known as kernel hyperparameters.
The squared-exponential kernel ensures that the surrogate predictions are smooth and differentiable with respect to the force field parameters, a property expected of ensemble structural features.

To improve surrogate accuracy for RDFs, we first subtract a physically motivated sigmoid baseline from the RDF training data of the form:

\begin{equation}
\label{eq:sigmoid}
    s(x) = \frac{1}{1 + e^{-a(x - x_0)}},
\end{equation}

\noindent with $x_0 = 3$~\AA{} and $a = 5$~\AA$^{-1}$.
This preprocessing step ensures the surrogate only learns deviations from the expected asymptotic RDF behavior and is equivalent to using Equation~\ref{eq:sigmoid} as the LGP prior mean~\cite{Sullivan2025-10.48550/arXiv.2507.07948}.

\textbf{Hyperparameter Learning:} The kernel hyperparameters for the $k$-th QoI, $\boldsymbol{\zeta_k}$, must be learned from the training data to ensure the accuracy and transferability of the LGP model. 
Here we construct a Bayesian hyperposterior from independent log-normal priors $\sim \mathcal{N}(-2, 2)$ over each parameter and a leave-one-out (LOO) log marginal likelihood (cf. Equation~8 in Reference~\cite{Sundararajan2001-10.1162/08997660151134343}):
\begin{align}
    \log p_\mathrm{LOO}(\mathbf{Y}^{(k)}|\boldsymbol{\zeta}_k) =
    & -\frac{1}{2N'\sigma_k^2} \sum_{i=1}^{N'} \frac{z^2(i)}{c_{ii}} \\
    & + \frac{\eta_k}{2N'} \sum_{i=1}^N \log c_{ii} - \frac{\eta_k}{2} \log(2\pi),
\end{align}
\noindent where $\mathbf{z} = \mathbf{C}^{-1} \mathbf{Y}^{(k)}$, $\mathbf{C} = \mathbf{K}^{(k)}_{\mathbf{X}, \mathbf{X}} + \sigma_k^2 \mathbf{I}$, and $c_{ii}$ denotes the $i$-th diagonal element of $\mathbf{C}^{-1}$.
Here, $N'$ is the number of training samples determined from an 80-20 train-test split of the full FFMD training set.
Minimization of the negative log hyperposterior to find the \textit{maximum a posteriori} is performed with stochastic gradient descent within the automatic differentiation pipeline in PyTorch~\cite{Paszke2019-}.
Performance and benchmarks of the LGP surrogates applied in this study are provided in Supplementary Section~S2.

In some applications, it is important to propagate uncertainty in the surrogate model predictions by marginalizing over the hyperparameter distributions of the hyperposterior.
If required, a Laplace approximation of the hyperposterior can be constructed as a multivariate normal distribution $p(\boldsymbol{\zeta}|\mathbf{y})\sim \mathcal{N}(\boldsymbol{\zeta}_\text{MAP}, \Sigma_\text{hyper})$ where $\Sigma_\text{hyper}$ is the inverse of the negative Hessian (which captures the curvature of the log-posterior around its mode).
Sampling from this Laplace approximation provides an efficient means to marginalize over hyperparameter uncertainty when needed. 
For our purposes, however, such marginalization was found to be unnecessary.
Results from a comparative test between a single LGP prediction and a 100 member committee average showed a negligible difference in the resulting force field parameter posteriors (Supplementary Figure~S4).

\paragraph{\textbf{Bayesian Inference of Force Field Parameters}}

Bayesian inference is a rigorous probabilistic approach to updating prior beliefs based on observed data. 
Its true power lies in the integration of physics-based expert priors to constrain model complexity, which naturally prevents overfitting while preserving principled uncertainty propagation through the parameter inference process.

According to Bayes' theorem, the posterior distribution of the model parameters $\boldsymbol{\theta}$ and nuisance parameters $\boldsymbol{n}$) given observed data $\boldsymbol{y}$ can be expressed as:

\begin{equation}
\label{eq:bayes}
p(\boldsymbol{\theta}, \boldsymbol{n}|\boldsymbol{y}) = \frac{p(\boldsymbol{y}|\boldsymbol{\theta}, \boldsymbol{n}) p(\boldsymbol{\theta}, \boldsymbol{n})}{p(\boldsymbol{y})},
\end{equation}

\noindent where $p(\boldsymbol{\theta}, \boldsymbol{n})$ is the joint prior reflecting our knowledge of $\boldsymbol{\theta}$ and $\boldsymbol{n}$ before observing the data, $p(\boldsymbol{y}|\boldsymbol{\theta}, \boldsymbol{n})$ is the likelihood of the data given the parameters, and the denominator $p(\boldsymbol{y})$ is the marginal likelihood (or model evidence), which serves as a normalization constant. We assume that the model parameters and nuisance parameters are independent in the prior so that $p(\boldsymbol{\theta}, \boldsymbol{n}) = p(\boldsymbol{\theta})p(\boldsymbol{n})$.

\textbf{Prior:} Each force field parameter $\boldsymbol{\theta}$ was assigned a weakly informative normal prior centered at the midpoint of the parameter range specified in data acquisition, with standard deviation set to $1/5$ of the range width.
This prior construction discourages sampling near boundaries where the surrogate model may be less reliable due to extrapolation error from edge effects.
One nuisance parameter per QoI is introduced to represent the combined effect of LGP uncertainty and observational variance in the likelihood model.
These are modeled as log-normally distributed (to avoid sampling negative variances) with $\log(\sigma)\sim \mathcal{N}(-2, 2)$.
We tested sensitivity to prior specification by comparing the normal priors to uniform (top-hat) priors spanning the same domain; posterior means and predictive distributions were essentially unchanged (Supplementary Figure~S2), indicating data-dominant posteriors for identified parameters.

\textbf{Likelihood:} We select a likelihood that assumes (i) QoIs are conditionally independent given the model parameters and (ii) residuals between predictions $\hat{\mathbf{y}}^{(k)}(\boldsymbol{\theta})$ and observations $\mathbf{y}^{(k)}$ are independent, homoskedastic, and Gaussian over the independent variables of each QoI:

\begin{equation}
    p(\boldsymbol{y}|\boldsymbol{\theta}, \boldsymbol{n})
    =
    \prod_{k=1}^K
    \frac{1}{\left(2\pi n_k^2\right)^{n^\mathrm{obs}_k/2}}
    \exp\left(
        -\frac{1}{2n_k^2}
        \left\| \hat{\mathbf{y}}_k(\boldsymbol{\theta}) - \mathbf{y}_k \right\|^2
    \right)
\end{equation}

\noindent where, for a given QoI $k$, there are $n^\mathrm{obs}$ independent observations and a nuisance parameter $n_{k}$.
In our implementation, we set the number of observations as one for each of the measured RDFs, hydrogen bond counts, and restrained distance distributions.
Assuming conditional independence in the likelihood greatly reduces the computational cost but neglects potential multivariate dependencies - for example, correlations between RDFs and hydrogen bond network metrics.
Likewise, the homoskedastic noise assumption reduces the number of inferred parameters but cannot capture known limiting behaviors of certain QoIs, such as the vanishing variance in RDFs at short range. 
More rigorous likelihoods that incorporate learned posterior covariances from reference data (cf. Reference~\citenum{Sullivan2025-10.48550/arXiv.2507.07948}) could be employed to relax these assumptions.

\textbf{Posterior:} We estimated the posterior distribution over $\boldsymbol{\theta}$ and $\boldsymbol{n}$ via Markov chain Monte Carlo sampling.
The number of walkers was set to $5\times$ the number of sampled dimensions (including nuisance parameters).
Initial walkers are drawn to satisfy bounds of the individual parameters as well as the total molecular charge constraint.
Any $\boldsymbol{\theta}$ violating parameter bounds or charge conservation constraint was assigned $p(\boldsymbol{\theta}) = 0$ and effectively rejected.
Walker moves were performed with the StretchMove algorithm with a maximum of 100,000 iterations.
Convergence was monitored using the integrated autocorrelation time $\tau$, with a minimum chain length of $100\tau$ and relative tolerance of $0.01$ as implemented in the \texttt{emcee} package~\cite{Foreman-Mackey2013-10.1086/670067, Goodman2010-10.2140/camcos.2010.5.65}.
The resulting chain was post processed by discarding the first $2\tau_{max}$ as a warm-up (or burn-in) and thinning the chain by an integer factor of $0.5\tau_{min}$, where $\tau_{max}$ and $\tau_{min}$ are the maximum and minimum integrated autocorrelation time over all inferred parameters, respectively.
The resulting posterior distribution delineates a clear and interpretable parameter landscape that enables estimation of optimal force field parameters and their associated uncertainties for a given system.

\textbf{Validation:} After the posterior distribution converged, ten charge parameter vectors $\boldsymbol{\theta}$ were sampled from a multivariate normal (Laplace) approximation of the posterior.
These samples were used to perform new FFMD simulations, following the same protocol as for the original training data acquisition.
The resulting QoI were then compared to reference AIMD data to evaluate the success of the learning procedure.

Agreement was quantified by the normalized mean absolute error (NMAE):
\begin{equation}
    \mathrm{NMAE}_k =
    \frac{\|\boldsymbol{y}_k - \widehat{\boldsymbol{y}}_k\|_1}{
    \|\boldsymbol{y}_k\|_1 + \|\widehat{\boldsymbol{y}}_k\|_1 },
\end{equation}
\noindent where $\|\cdot\|_1 := \sum_{i=1}^{\eta_k}|\cdot_i|$ denotes the L$^1$ norm.
By construction, $\mathrm{NMAE}=0$ indicates perfect agreement and $\mathrm{NMAE}=1$ indicates complete disagreement. For clarity, we report NMAE as a percentage.

\subsection{\textit{Ab initio} molecular dynamics}
\textit{Ab initio} molecular dynamics (AIMD) simulations were conducted using the CP2K~9.1~\cite{Kuhne2020-10.1063/5.0007045a} software package, using the QUICKSTEP module for density functional theory (DFT) calculations.
The electronic ground-state energies and forces were described under the Born--Oppenheimer approximation using the revPBE~\cite{Perdew1996-10.1103/PhysRevLett.77.3865a, Zhang1998-10.1103/PhysRevLett.80.890a} exchange-correlation functional, incorporating the D3 dispersion correction~\cite{Grimme2010-10.1063/1.3382344}, with exclusions applied to Ca$^{2+}$-containing pairs~\cite{Kostal2023-10.1021/acs.jpclett.3c00856a}.
Kohn-Sham orbitals were represented using Gaussian TZV2P basis sets~\cite{VandeVondele2007-10.1063/1.2770708a} in conjunction with Godecker-Hutter-Tetter (GTH) pseudopotentials~\cite{Goedecker1996-10.1103/PhysRevB.54.1703a} for core electrons, and the electronic density was expanded in plane waves with a 400~Ry cutoff.
The simulations were performed in the NVT ensemble at 300~K.
An initial 5~ps equilibration phase was carried out using a Langevin thermostat with a friction coefficient of 0.02.
Subsequently, the production runs employed a stochastic velocity rescaling thermostat (SVR)~\cite{Bussi2007-10.1063/1.2408420a} with a time constant of 1~ps.
The equations of motion were integrated using the velocity Verlet algorithm with a time step of 0.5~fs.
An overview of the composition and simulation details for species is provided in Supplementary Table~S3.

\subsection{Force field molecular dynamics}

All the FFMD simulations were carried out in Gromacs~2024.3~\cite{Abraham2015-10.1016/j.softx.2015.06.001a}.

\textbf{training set simulations:} The molecular force fields were taken as from CHARMM-GUI~\cite{Lee2016-10.1021/acs.jctc.5b00935}.
Each species was initially solvated in a cubic box of roughly 1.6~nm containing 128 TIP4P/2005 water molecules~\cite{Abascal2005-10.1063/1.2121687}, optionally with counterions added.
The system was first energy-minimized using the steepest descent algorithm.
Subsequently, a 10~ns NpT equilibration was performed (discarding the first 1~ns) to determine the average box size.
Pressure was maintained at 1~bar using the C-rescale~\cite{Bernetti2020-10.1063/5.0020514} barostat with a time constant of 1~ps and a compressibility of $4.5 \cdot 10^{-5}$~bar$^{-1}$.
Temperature was maintained at 300~K using SVR thermostat with a 1~ps time constant, followed by a 1~ns NVT equilibration using the same thermostat.
Hydrogen bond constraints were applied via the LINCS~\cite{Hess1997-10.1002/SICI1096-987X19970918:12<1463::AID-JCC4>3.0.CO;2-H} algorithm.
Nonbonded interactions were computed using a cutoff scheme with a 0.7~nm cutoff and long-range van der Waals and Coulombic interactions were treated using particle mesh Ewald (PME)~\cite{Essmann1995-10.1063/1.470117} method with a potential-shift modifier.
Such equilibrated structures were then used as initial conditions for both the training set generation and the reference AIMD simulations.

\textbf{density simulations:}
For bulk solution density calculations, cubic boxes (4.5nm per side) containing 2000 water molecules and the appropriate number of solutes were constructed to match target concentrations.
Each system underwent steepest-descent minimization, 200ps of NpT equilibration, and a 1.2ns production run in the NpT ensemble at 300K and 1bar.
The SVR thermostat (1ps) and C-rescale barostat (5~ps, $4.5\times10^{-5}$bar$^{-1}$) were used for temperature and pressure control, respectively.
Nonbonded cutoffs were set to 1.2nm, with all other parameters identical to those in the training set simulations.

Experimental densities were obtained from Reference~\citenum{Haynes2016-10.1201/9781315380476a} for aqueous sodium acetate, ethanol, and ammonium chloride; from Reference~\citenum{Sohnel1985-} for potassium hydrogenphosphate, potassium dihydrogenphosphate, and potassium formate; from Reference~\citenum{Kumar2001-10.1023/A:1005231617292} for guanidinium chloride; and from Reference~\citenum{Assarsson2002-10.1021/j100854a004} for \textit{N}-acetamide.
To compare simulated and experimental data at matching concentrations, experimental density data were fitted with a second-order polynomial.

\textbf{Troponin simulations:}
Initial coordinates for N-cTnC were taken from the crystal structure (PDB ID: 1AP4~\cite{Spyracopoulos1997-10.1021/bi971223d}, neutralized with K$^+$ ions, and solvated in a 5.25nm cubic box containing 150mM KCl using CHARMM-GUI~\cite{Lee2016-10.1021/acs.jctc.5b00935}.
Each system underwent energy minimization, followed by 5ns NpT equilibration.

For simulations involving charge sets developed here, water and ions compatible with the ECC were employed~\cite{CrucesChamorro2024-10.1021/acs.jpclett.4c00344, Fan2025-10.1021/acs.jctc.5c00873}.
Nonbonded interactions were computed the same way as in the training set simulations with a 1.2~nm cutoff.
In the case of CHARMM36, CHARMM36-nbfix and ProsECCo75, long-range electrostatic interactions were computed using the PME method while vdW interactions were truncated and smoothened by a force-switch modifier.

All the simulations under the NpT ensemble were performed at 310~K maintained by the SVR thermostat with 1~ps time constant and pressure was controlled by C-rescale barostat at 1~bar using 5~ps time constant and isotropic compressibility of $4.5\times10^{-5}$bar$^{-1}$~bar$^{-1}$.

\subsection{Alchemical binding free energy calculation}

\begin{figure}
    \centering
    \includegraphics[width=0.8\linewidth]{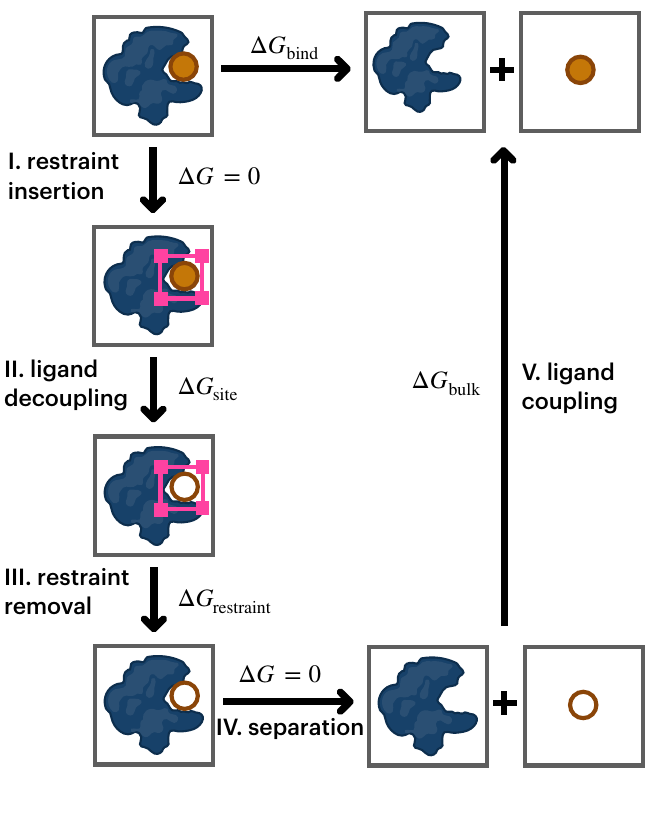}
    \caption
    {
    The thermodynamic cycle used to calculate the standard binding free energy of Ca$^{2+}$ to the EF-hand loop of the regulatory domain of troponin.
    The protein is shown in blue, Ca$^{2+}$ in orange, and restraints in pink.
    A filled orange circle indicates Ca$^{2+}$ fully interacting (coupled) with its environment, while an open circle denotes the decoupled state.
    The ink square represents the imposed constraints.
    }    
    \label{fig:alchemical-free-energy}
\end{figure}

The binding free energy of a Ca$^{2+}$ ion to the EF-hand loop of troponin was calculated using the alchemical double decoupling technique~\cite{Salari2018-10.1021/acs.jctc.8b00447, Duboue-Dijon2021-10.1063/5.0046853}, as illustrated in the thermodynamic cycle in Figure~\ref{fig:alchemical-free-energy}.
The binding site was defined by the distances between Ca$^{2+}$ and its coordinating atoms for each force field (see Supplementary Figure~S7).
The equilibrium geometry of the site was obtained from a 100ns NpT simulation.
Harmonic flat-bottom restraints were applied between Ca$^{2+}$ and its coordinating atoms to prevent ion dissociation during the decoupling process.
Because the binding site remained stable throughout the simulation, these restraints were assumed to contribute no free energy in the fully coupled state (stage~I), as they did not perturb the natural fluctuations of the bound complex.
The corresponding equilibrium geometries and restraint parameters are listed in Supplementary Table~S2.
Two independent alchemical transformations were carried out.
In the first, the ligand was gradually decoupled from its environment within the binding site of the protein (stage~II); in the second, the ligand was decoupled from bulk solvent (stage~V).
Each transformation comprised 21 $\lambda$-windows.
During decoupling, electrostatic interactions were turned off first over 11 windows, followed by van der Waals interactions over 10 windows, with $\lambda$ incremented in steps of $0.1$.
Each window was simulated for 100~ns during stage I and 10~ns during stage V. 

Free energy differences between adjacent windows were computed using the Bennett acceptance ratio (BAR) as implemented in Gromacs using \texttt{gmx bar} command while discarding the first 1~ns for equilibration.
Both bulk and site decoupling calculations were corrected for the neutralizing background charge introduced by the PME when the net system charge changed during Coulombic decoupling (Equation~17 in Referece~\citenum{Simonson2016-10.1080/08927022.2015.1121544}):
\begin{equation}
\label{eq:pbc-correction}
    \Delta G^\mathrm{pbc} = \frac{|\zeta|(q_\mathrm{out}^2-q_\mathrm{in}^2)}{2\varepsilon L},
\end{equation}
where $\zeta=-2.837$ is a constant, $\varepsilon=80$ is the is the dielectric constant of the solvent, $L$ is the length of the simulation box, and $q_\mathrm{in}$, $q_\mathrm{out}$ are the net charges of the initial and final states, respectively.

To prevent the ion to escape from the binding pocket during decoupling, harmonic flat-bottom distance restraints were applied between the ligand and coordinating atoms.
It was assumed that the introduction of restraints contributes no free energy in the fully coupled state (stage I), as the binding site is stable and the restraint does not perturb natural ligand fluctuations.
The free energy associated with releasing the restraints in the decoupled state (stage III) was evaluated analytically for a single flat-bottom potential and by thermodynamic integration for the rest.
The restraint correction accounts for the reduction of configurational volume accessible to the ligand:
\begin{equation}
    \label{eq:deltaG_restraint}
    \Delta G_\mathrm{restraint}=-RT\ln \left( \frac{V}{V^0} \right),
\end{equation}
where $V^0$ is the standard volume occupied by one molecule at concentration of $1~\mathrm{mol}/\mathrm{dm}^3$ and $V$ is the effective binding site volume defined by the restraint potential $U(r)$:
\begin{equation}
\label{eq:standard-volume}
    V_0=\frac{1\mathrm{dm}^3}{N_A}=\frac{10^{24}\mathrm{nm}^3}{6.022\cdot 10^{23}}=1.661\mathrm{nm}^3
\end{equation}
\begin{equation}
\label{eq:restraint-volume}
    V=\int_0^\infty \mathrm{d}r\ 4\pi r^2 e^{-\beta U}
\end{equation}
For a single flat-bottom restraint the integral in Equation~\ref{eq:restraint-volume} is analytical; for additional restraints the free energy from Equation~\ref{eq:deltaG_restraint} was estimated using thermodynamic integration by gradually reducing the restraint force constant according to $k_\lambda=k_0 + \lambda^\alpha(k_1-k_0)$, with $\alpha=4$ chosen to ensure smooth convergence near $k_\lambda=0$.
Each $\lambda$-window was simulated for 22~ns, discarding 2~ns due to equilibration.
All restraints were implemented and removed using the \texttt{Colvars} module~\cite{Fiorin2013-10.1080/00268976.2013.813594, Fiorin2024-10.1021/acs.jpcb.4c05604}.
A second assumption is made that the free energy cost of partitioning the system into two independent simulation boxes is zero once the ligand is fully decoupled and no restraints remain (stage IV in Figure~\ref{fig:alchemical-free-energy}).

The overall standard binding free energy was then obtained as:
\begin{equation}
    \Delta G_\mathrm{bind} = \Delta G_\mathrm{bulk} + \Delta G_\mathrm{site} + \Delta G_\mathrm{restraint}
\end{equation}

\section*{Data Availability}
Open source code will be uploaded pending revisions, and may be available from the authors upon reasonable request.

\section*{Acknowledgements}
P.J. acknowledges support from an ERC Advanced Grant (grant agreement no. 101095957).
V.K. acknowledges support from the Charles University where he is enrolled as a PhD. student and from the IMPRS-QDC Dresden.
We thank Elise Duboué-Dijon and Jérôme Hénin for help with the alchemical decoupling setup.

\section*{References}

\end{document}


\def\mstitle{Supplementary Information: Bayesian learning for accurate and robust biomolecular force fields}

\title{\mstitle}

\author{Vojtech Kostal}

\author{Brennon L. Shanks}

\author{Pavel Jungwirth*}
\email{pavel.jungwirth@uochb.cas.cz}

\author{Hector Martinez-Seara*}
\email{hseara@gmail.com}
\affiliation{
Institute of Organic Chemistry and Biochemistry of the Czech Academy of Sciences, Flemingovo nám. 2, 166 10 Prague 6, Czech Republic
}

\date{\today}
\maketitle

\section{The Probability Model}

This section provides the full probabilistic formulation underlying the Bayesian inference of FFMD parameters from AIMD reference data, including surrogate marginalization and hyperparameter uncertainty. This formalism underpins the simplified expressions used in the main text.

\subsection{The Full Probability Model}

The full joint probability model is symbolically written as:
\begin{equation}
    p(\mathcal{D}, \boldsymbol{\theta}, \boldsymbol{\zeta}, \mathcal{S}_{\boldsymbol{\zeta}}, \mathcal{Y})
\end{equation}
\noindent where $\mathcal{D}$ is the FFMD training dataset, $\boldsymbol{\theta}$ is the set of FFMD force field parameters, $\boldsymbol{\zeta}$ is the set of surrogate model hyperparameters, $\mathcal{S}_{\boldsymbol{\zeta}}$ is the LGP surrogate model, and $\mathcal{Y}$ is the AIMD simulation reference set. Note that we have removed the subscript $k$ notation specifying a specific QoI for convenience.

In principle, defining this full joint distribution enables exact Bayesian inference over all model variables. Specifically, by marginalizing and conditioning appropriately, one can obtain posterior estimates of the optimal parameter set $\boldsymbol{\theta}$ along with rigorous uncertainty quantification.

The first approximation we make is that the FFMD training set $\mathcal{D}$ is observed and fixed; we therefore treat it as given and do not place a prior distribution over it. The full probability model is then:
\begin{equation}
     p(\mathcal{D}, \boldsymbol{\theta}, \boldsymbol{\zeta}, \mathcal{S}_{\boldsymbol{\zeta}}, \mathcal{Y}) =  p(\boldsymbol{\theta}, \boldsymbol{\zeta}, \mathcal{S}_{\boldsymbol{\zeta}}, \mathcal{Y} \mid \mathcal{D})
\end{equation}
\noindent which forms the basis for all subsequent inference. 

The choice of factorization of the joint probability model reflects our assumptions about the hierarchy of model inputs, defining a generative process. Here our model assumes that the hyperparameters give rise to surrogate models, which in turn define distributions over observables conditioned on the force field parameters. The factorization reflecting this hierarchy is:
\begin{align}
    p(\boldsymbol{\theta}, \boldsymbol{\zeta}, \mathcal{S}_{\boldsymbol{\zeta}}, \mathcal{Y} \mid \mathcal{D}) 
    &= \, p(\boldsymbol{\zeta} \mid \mathcal{D}) \notag \\
    &\quad \times p(\mathcal{S}_{\boldsymbol{\zeta}} \mid \boldsymbol{\zeta}, \mathcal{D}) \notag \\
    &\quad \times p(\boldsymbol{\theta} \mid \mathcal{S}_{\boldsymbol{\zeta}}, \boldsymbol{\zeta}, \mathcal{D}) \notag \\
    &\quad \times p(\mathcal{Y} \mid \boldsymbol{\theta}, \mathcal{S}_{\boldsymbol{\zeta}}, \boldsymbol{\zeta}, \mathcal{D})
\end{align}

Now, we will assume that the probability distribution $p(\boldsymbol{\theta} \mid \mathcal{S}_{\boldsymbol{\zeta}}, \boldsymbol{\zeta}, \mathcal{D})$ is conditionally independent of $(\mathcal{S}_{\boldsymbol{\zeta}}, \boldsymbol{\zeta}, \mathcal{D})$:
\begin{equation}
    p(\boldsymbol{\theta} \mid \mathcal{S}_{\boldsymbol{\zeta}}, \boldsymbol{\zeta}, \mathcal{D}) = p(\boldsymbol{\theta})
\end{equation}
\noindent In other words, we place a prior directly on $\boldsymbol{\theta}$ that does not depend on the surrogate model, its hyperparameters, or the training data. 

The second simplifying assumption is that the likelihood $p(\mathcal{Y} \mid \boldsymbol{\theta}, \mathcal{S}_{\boldsymbol{\zeta}}, \boldsymbol{\zeta}, \mathcal{D})$ depends on its inputs only through the evaluation of the surrogate model at parameter value $\boldsymbol{\theta}$:
\begin{equation}
    p(\mathcal{Y} \mid \boldsymbol{\theta}, \mathcal{S}_{\boldsymbol{\zeta}}, \boldsymbol{\zeta}, \mathcal{D}) = p(\mathcal{Y} \mid \mathcal{S}_{\boldsymbol{\zeta}}(\boldsymbol{\theta}))
\end{equation}
\noindent In other words, the surrogate model acts as a probabilistic emulator of the underlying physical model, and all downstream predictions are mediated through it.

The probability model then becomes:
\begin{align}
    p(\boldsymbol{\theta}, \boldsymbol{\zeta}, \mathcal{S}_{\boldsymbol{\zeta}}, \mathcal{Y} \mid \mathcal{D}) 
    &= 
    \overbrace{p(\boldsymbol{\zeta} \mid \mathcal{D})}^{\text{Hyperposterior}} \notag \\
    &\quad \times \overbrace{p(\mathcal{S}_{\boldsymbol{\zeta}} \mid \boldsymbol{\zeta}, \mathcal{D})}^{\text{Surrogate Distribution}} \notag \\
    &\quad \times \overbrace{p(\boldsymbol{\theta})}^{\text{Parameter Prior}} \notag \\
    &\quad \times \overbrace{p(\mathcal{Y} \mid \mathcal{S}_{\boldsymbol{\zeta}}(\boldsymbol{\theta}))}^{\text{Likelihood}}
\end{align}

These factorizations significantly simplify the inference process by reducing the dimensionality of integration and allowing sequential treatment of components, such as learning the hyperposterior independently from downstream parameter inference.

\subsection{Learning the Parameter Posterior}

Our aim is to compute the posterior distribution of FFMD model parameters given FFMD training data and an AIMD reference, which can be written using Bayes' theorem:
\begin{equation}
    p(\boldsymbol{\theta}\mid\mathcal{D}, \mathcal{Y}) = \frac{p(\mathcal{D}, \mathcal{Y}\mid\boldsymbol{\theta})p(\boldsymbol{\theta})}{p(\mathcal{D}, \mathcal{Y})}
\end{equation}
\noindent which, upon factorizing the numerator and denominator gives:
\begin{equation}
\frac{p(\mathcal{Y}\mid\boldsymbol{\theta},\mathcal{D})p(\mathcal{D}\mid\boldsymbol{\theta})p(\boldsymbol{\theta})}{p(\mathcal{Y}\mid\mathcal{D})p(\mathcal{D})}
\end{equation}
\noindent But we assumed $\mathcal{D}$ and $\boldsymbol{\theta}$ are conditionally independent ($p(\mathcal{D}\mid\boldsymbol{\theta}) = p(\mathcal{D})$), so we obtain:
\begin{equation}
    \frac{p(\mathcal{Y}\mid\boldsymbol{\theta},\mathcal{D})p(\boldsymbol{\theta})}{p(\mathcal{Y}\mid\mathcal{D})} \propto p(\mathcal{Y}\mid\boldsymbol{\theta},\mathcal{D})p(\boldsymbol{\theta})
\end{equation}
\noindent indicating that we just need to specify a prior over $\boldsymbol{\theta}$ and a likelihood. The way we evaluate the likelihood will depend on whether we explicitly learn the hyperposterior as in our full joint model or if we take a point estimate like in MAP estimation.

\subsubsection{Case 1: Learning the Hyperposterior}

If we want to treat the problem in a fully Bayesian way, uncertainty propagation from the hyperparameters and surrogate model should be accounted for explicitly. One can see this by rewriting the likelihood as the joint marginal of full probability model conditioned on $\boldsymbol{\theta}$ with respect to $\boldsymbol{\zeta}$ and $\mathcal{S}_{\boldsymbol{\zeta}}$:
\begin{equation}
    p(\mathcal{Y}\mid\boldsymbol{\theta},\mathcal{D}) = \int d\boldsymbol{\zeta} \int d\mathcal{S}_{\boldsymbol{\zeta}} \space p(\boldsymbol{\zeta}, \mathcal{S}_{\boldsymbol{\zeta}}, \mathcal{Y} \mid \boldsymbol{\theta}, \mathcal{D})
\end{equation}
\noindent where the integrals are understood to go over the domain of each input. Writing this out explicitly reveals an inner and outer integral:
\begin{equation}
    =\underbrace{\int p(\boldsymbol{\zeta} \mid \mathcal{D}) \,
    \overbrace{\left[ \int p(\mathcal{S}_{\boldsymbol{\zeta}} \mid \boldsymbol{\zeta}, \mathcal{D}) \,
    p\big(\mathcal{Y} \mid \mathcal{S}_{\boldsymbol{\zeta}}(\boldsymbol{\theta})\big) \, d\mathcal{S}_{\boldsymbol{\zeta}} \right]}^{\text{Expected Likelihood under GP Posterior}} \,
    d\boldsymbol{\zeta}}_{\text{Integration over Hyperposterior}}
\end{equation}

\noindent In practice, we can approximate the likelihood by discrete Monte Carlo sampling so that:
\begin{align}
    &\int p(\boldsymbol{\zeta} \mid \mathcal{D}) \left[ 
        \int p(\mathcal{S}_{\boldsymbol{\zeta}} \mid \boldsymbol{\zeta}, \mathcal{D}) \, 
        p\big(\mathcal{Y} \mid \mathcal{S}_{\boldsymbol{\zeta}}(\boldsymbol{\theta})\big) \, 
        d\mathcal{S}_{\boldsymbol{\zeta}} 
    \right] d\boldsymbol{\zeta} \nonumber \\
    &\quad \approx 
    \frac{1}{N} \sum_{i=1}^{N} \left[
        \frac{1}{M} \sum_{j=1}^{M} 
        p\big( \mathcal{Y} \mid \mathcal{S}^{(j)}_{\boldsymbol{\zeta}^{(i)}}(\boldsymbol{\theta}) \big)
    \right]
\end{align}
\noindent which indicates that we compute a double sum of the likelihood over samples of the GP posterior with a sample drawn from the hyperposterior. If the surrogate model variance is negligibly small, we can further approximate the surrogate predictive distribution as a delta function centered on the surrogate posterior mean so that:

\begin{equation}
    \boldsymbol{\mu}_{\boldsymbol{\zeta}}(\boldsymbol{\theta})
    = \mathbb{E}_{p(\mathcal{S}_{\boldsymbol{\zeta}}\mid\boldsymbol{\zeta},\mathcal{D})}
    \!\big[\mathcal{S}_{\boldsymbol{\zeta}}(\boldsymbol{\theta})\big],
\end{equation}

\noindent which causes the inner integral to simplify as:

\begin{align}
\label{eq:approx_hyperposterior}
 \int p(\boldsymbol{\zeta} \mid \mathcal{D}) 
    p\!\left( \mathcal{Y} \mid \boldsymbol{\mu}_{\boldsymbol{\zeta}}(\boldsymbol{\theta}) \right) d\boldsymbol{\zeta}
    &\approx \frac{1}{N} \sum_{i=1}^N 
    p\!\left( \mathcal{Y} \mid \boldsymbol{\mu}_{\boldsymbol{\zeta}^{(i)}}(\boldsymbol{\theta}) \right).
\end{align}

\noindent This assumption effectively states that epistemic uncertainty in the surrogate predictions is negligible.

\subsubsection{Case 2: MAP Estimation of the Hyperposterior}

In \textit{maximum a posteriori} (MAP) estimation of the hyperposterior, we retain only the most probable hyperparameter set and treat it as fixed. 
This is equivalent to approximating the hyperposterior as a delta function centered at its MAP estimate $p(\boldsymbol{\zeta} \mid \mathcal{D}) \approx \delta(\boldsymbol{\zeta} - \boldsymbol{\zeta}_{\text{MAP}})$, which further simplifies the likelihood:
\begin{equation}
    \int p(\boldsymbol{\zeta} \mid \mathcal{D}) \, 
    p\left( \mathcal{Y} \mid \boldsymbol{\mu}_{\boldsymbol{\zeta}}(\boldsymbol{\theta}) \right) \, d\boldsymbol{\zeta}
    \approx p\left( \mathcal{Y} \mid \boldsymbol{\mu}_{\boldsymbol{\zeta}_{\text{MAP}}}(\boldsymbol{\theta}) \right)
\end{equation}
where the delta-function approximation allows us to bypass marginalization over surrogate hyperparameters. 
The method presented is capable of performing any of these tasks depending on the user-defined goals and degree of care placed on the uncertainty quantification step.

\section{Local Gaussian Process benchmarks}

\begin{figure*}
    \centering
    \begin{minipage}[t]{0.45\textwidth}
        \centering
        \includegraphics[width=\linewidth]{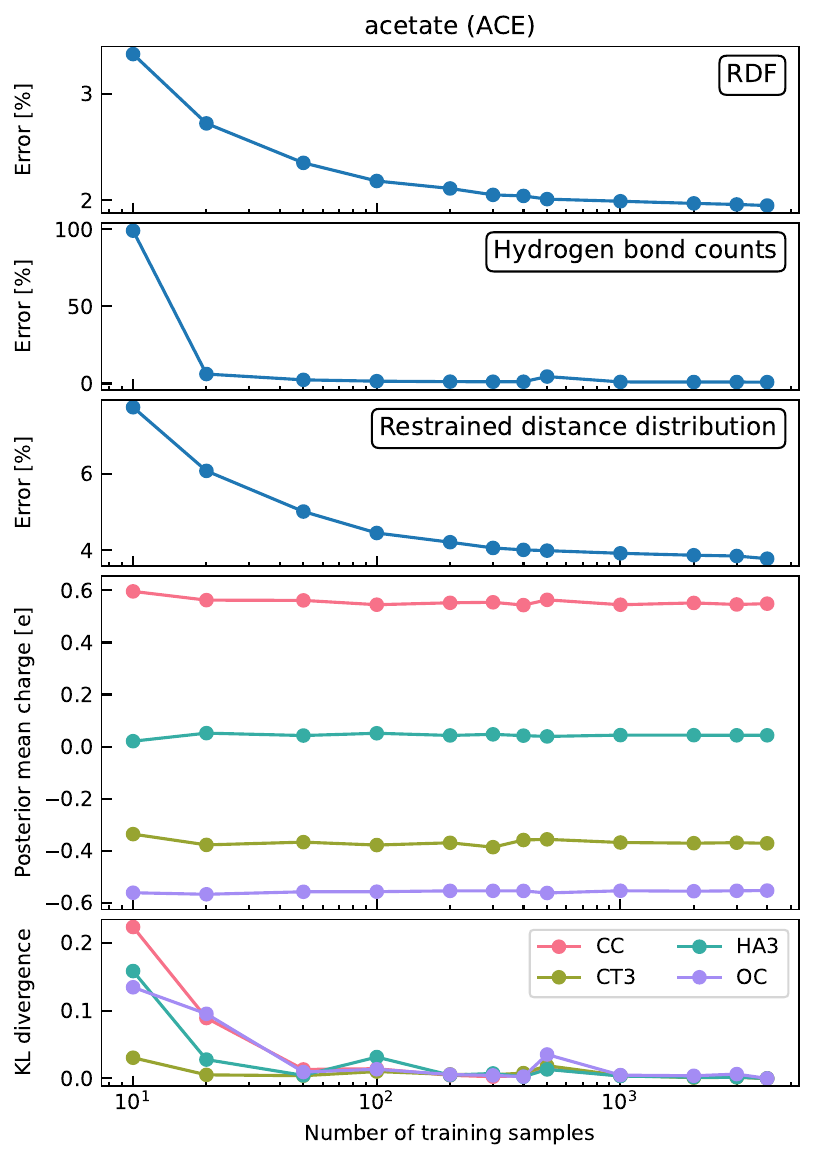}
        \par\small (a)
    \end{minipage}
    \hfill
    \begin{minipage}[t]{0.45\textwidth}
        \centering
        \includegraphics[width=\linewidth]{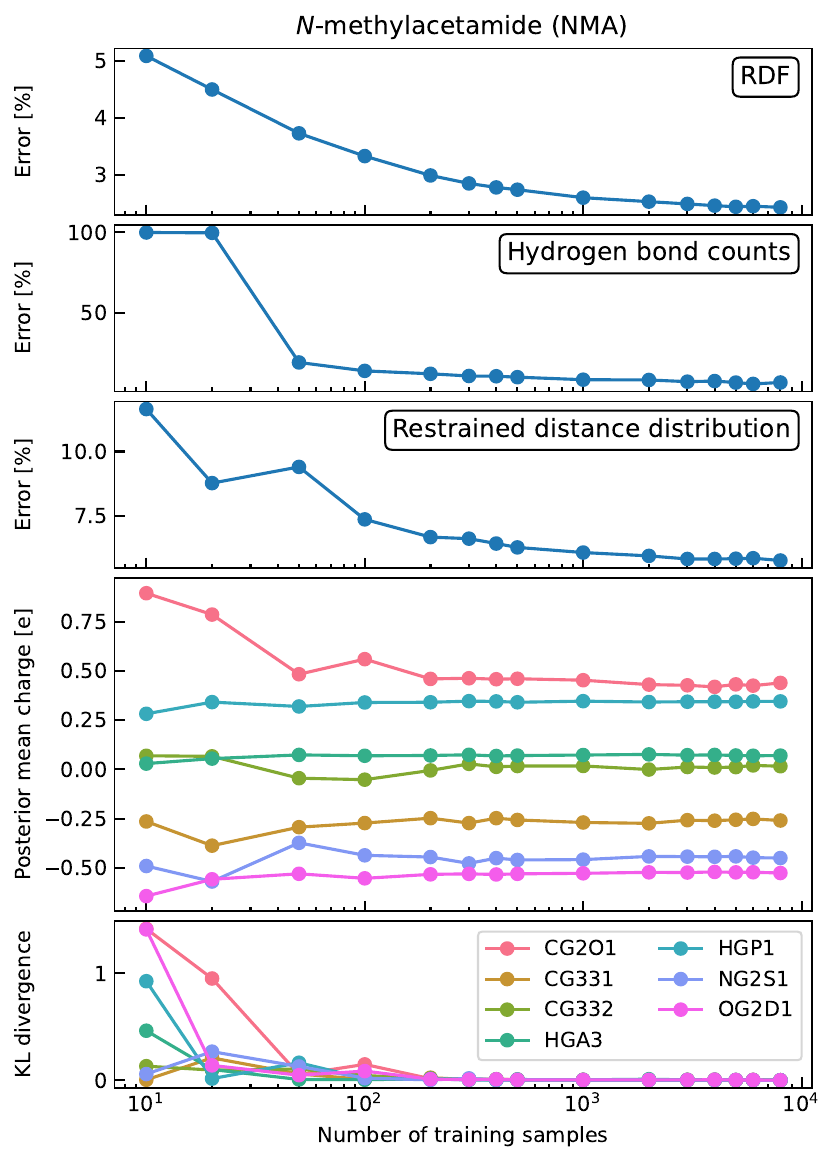}
        \par\small (b)
    \end{minipage}
    \caption{Performance of the LGP surrogate models and optimization results as functions of the training set size for acetate (left) and \textit{N}-methylacetamide (right).
    The first three panels show prediction errors for individual quantities of interest (QoIs).
    The fourth panel shows the convergence of the posterior means, and the fifth shows the Kullback--Leibler divergence between posteriors at different training sizes and the final posterior.}
    \label{fig:benchmarks}
\end{figure*}

Figure~\ref{fig:benchmarks} shows benchmark results for the optimization procedure applied to acetate and \textit{N}-methylacetamide that were chosen as representative systems with lower (3) and higher (6) numbers of optimized partial charges, respectively.
For each molecule, we repeated the full optimization process independently for varying training set sizes.
The number of training samples was controlled by adjusting the train/test split fraction.
At each training size, hyperparameters of the local Gaussian process (LGP) surrogate models—one per quantity of interest (QoI)—were optimized using up to 200 training samples, i.e., $\min(n_\mathrm{train}, 200)$.
Using these optimized hyperparameters, we inferred posterior distributions over the parameters via the Bayesian inference scheme described in the main text, employing MCMC sampling.

The first three panels of Figure~\ref{fig:benchmarks} show the prediction errors over the test set of the LGP surrogates for individual QoI as a function of the number of training samples.
The fourth panel displays the convergence of the most probable parameter values, defined as the mode of the posterior distributions.
The fifth panel shows the Kullback--Leibler (KL) divergence between the posterior obtained at a given training size and the reference posterior inferred from the largest training set.

The total number of available samples was 5000 for acetate and 10000 for \textit{N}-methylacetamide.
A maximum test set fraction of 0.2 was used, resulting in up to 4000 and 8000 training samples, respectively.

Across both systems, all reported metrics stabilize between 500 and 1000 training samples, indicating that this range is sufficient for reliable optimization.
Notably, despite the difference in dimensionality of the optimization problems (3 vs. 6 parameters), the number of required training samples to achieve convergence is comparable.

\section{Bayesian learning benchmarks}

Bayesian inference represents the gold standard approach for parameter uncertainty quantification, but important diagnostic tests are important to ensure that the inference is robust \cite{VanDeSchoot2021-10.1038/s43586-020-00001-2}. 
The most important are evaluations are prior sensitivity, likelihood sensitivity to the QoI, and the posterior sensitivity to hyperparameter uncertainty.
Here, we perform diagnostic tests for all three cases to evaluate the robustness of our predicted posteriors.

\subsection{Posterior Sensitivity to Prior - Informative vs Diffuse}

\begin{figure}
    \centering
    \includegraphics[width=\linewidth]{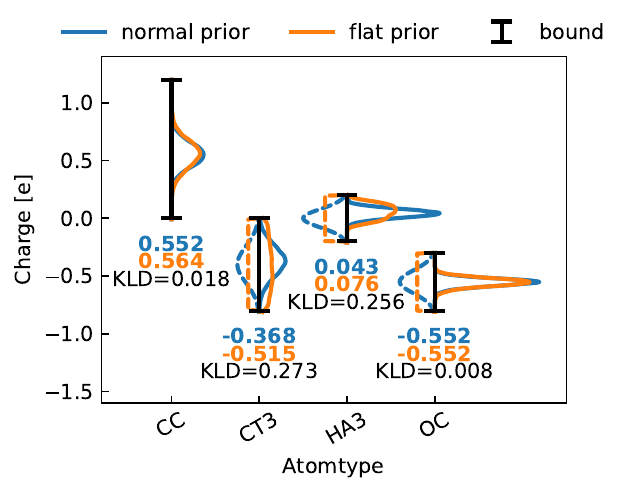}
    \caption{Optimization results for acetate using a flat (orange) or normal (blue) prior.
    The x-axis lists atom types, the y-axis their partial charges.
    Dashed lines show priors, solid lines posteriors; vertical capped black lines mark allowed bounds.
    The implicit atom type has no prior.
    Posterior modes are shown below each bound, color-coded by prior type, and black numbers indicate the Kullback–Leibler divergence (KLD) between posteriors from the two priors.}
    \label{fig:flat-normal-prior}
\end{figure}

Figure~\ref{fig:flat-normal-prior} shows the converged posterior distributions obtained using two different prior types: a diffuse flat prior (uniform, orange) and a weakly informative normal prior (Gaussian, blue) prior (right) plotted as dashed lines.
This comparison serves to assess the sensitivity of the optimization procedure to the choice of prior distribution.
The flat prior represents a diffuse prior assumption, placing equal probability across the entire allowed range.
In contrast, the normal prior is centered in the middle of the parameter bounds, reflecting the expectation that these bounds were chosen based on typical charge values found in existing force fields.
To quantify the similarity between the posterior distributions obtained from each prior type, we compute the Kullback--Leibler (KL) divergence, which is shown in black under each atom type result in Figure~\ref{fig:flat-normal-prior}.

For some atom types, such as OC and CC, the posterior distributions are nearly identical regardless of the prior used, as indicated by small KL divergence values.
This implies that for these parameters, the data are informative enough to override prior assumptions and posterior converges to the same distribution regardless of the prior shape.
In contrast, atom types such as CT3 and HA3 show larger divergence between posteriors, with the posterior closely resembling the prior in both cases.
This indicates that the quantities of interest (QoIs) used in the optimization are not sensitive to these parameters.
As a result, the optimization cannot constrain their values meaningfully, and any value within the allowed range could fulfill the reference data equally well.
For parameter HA3, the posterior shape does reflect the prior: when using a normal prior, the posterior tends to be narrower, a consequence of the normal prior assigning lower probabilities near the boundaries. 

Overall, the normal prior is recommended.
It produces posteriors that are centered well within the parameter bounds, a reasonable choice given the lack of specific knowledge about optimal charge values.
Although the posterior shape varies somewhat with different priors, the most probable values (color-coded in Figure \ref{fig:flat-normal-prior}) are largely consistent.
The main exception is the CT3 atom type, where the flat prior yields a nearly uniform posterior; its mode then becomes sensitive to small numerical fluctuations, which are not physically meaningful.
Moreover, surrogate models tend to be less accurate near parameter boundaries, so a prior that naturally downweights these regions enhances the robustness of the inference.

\subsection{Posterior Sensitivity to QoI Selection}

\begin{figure}[b]
    \centering
    \includegraphics[width=\linewidth]{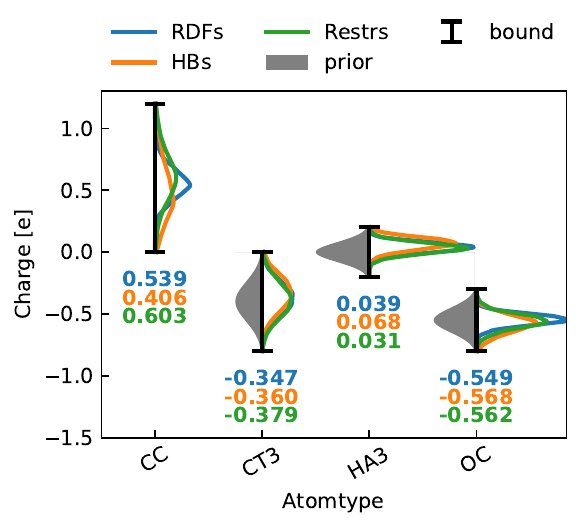}
    \caption{Partial charge learning results for acetate using using the QoIs separately.}
    \label{fig:post-qoi}
\end{figure}

We benchmarked the influence of three QoIs -- radial distribution functions (RDFs), average hydrogen bond counts (HBs), and restrained distance distributions (Restrs) -- on the final posterior.
To this end, we optimized the partial charges of acetate considering only one QoI at a time.
The resulting posteriors (Figure \ref{fig:post-qoi}) from these independent optimizations are highly consistent, with differences being primarily quantitative rather than qualitative.
When all three QoIs are incorporated simultaneously, the optimization yields parameter values that balance the individual optima, yet still fall well within one standard deviation of the respective posterior distributions --indicating strong agreement and mutual compatibility between the QoI.

\subsection{Posterior Sensitivity to Hyperposterior Marginalization}

\begin{figure}
    \centering
    \includegraphics[width=0.8\linewidth]{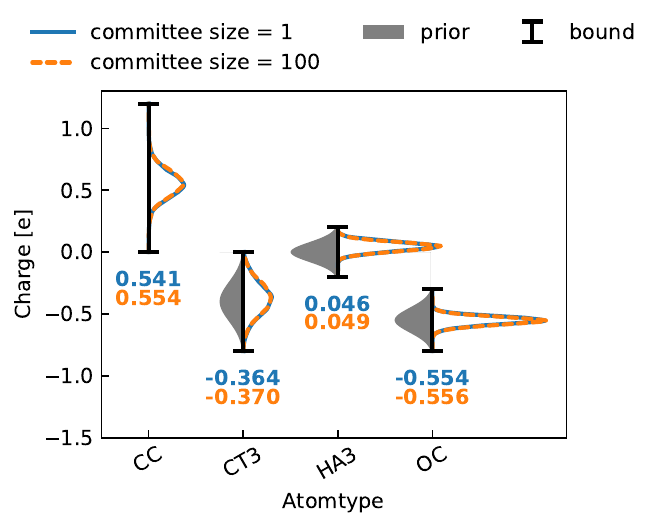}
    \caption{Partial charge learning results for acetate using using either 1 (blue) or 100 (orange) members of the LGP committee.}
    \label{fig:committee-size}
\end{figure}

In Figure~\ref{fig:committee-size}, we examine the effect of LGP committee size on the optimization results for acetate.
The solid blue line shows the posterior obtained using a single surrogate model to predict the QoIs with optimal hyperparameters.
The dashed orange line corresponds to results from an ensemble average of a committee of 100 LGPs according to equation \eqref{eq:approx_hyperposterior}, each with hyperparameters drawn from a Laplace approximation of the hyperposterior.
This ensemble averaging approximates the influence of hyperparameter uncertainty on the posterior distribution.
Notably, the resulting posterior differs only marginally from that obtained using a single model with the MAP hyperparameters, supporting the practical use of the MAP estimate alone.

\newpage

\section{Comparison with CHARMM36-nbfix}

\begin{figure}
    \centering
    \includegraphics[width=\linewidth]{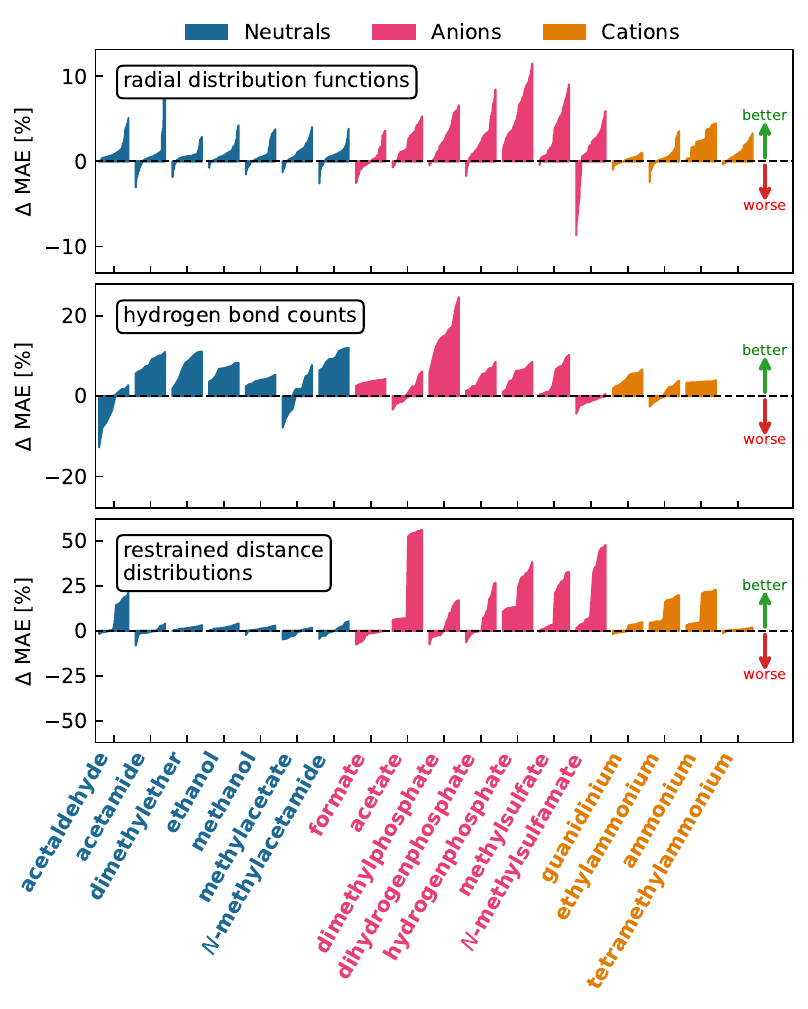}
    \caption{
        Silhouette plots capturing the absolute differences between the NMAE score obtained by the original force field and the force field with optimized partial charges for the molecules on the x-axis.
        Molecules are grouped and color-coded by chemical class: neutrals (purple), anions (red), and cations (gold).
    }
    \label{fig:fc_vs_ecc80}
\end{figure}

Figure~\ref{fig:fc_vs_ecc80} shows the absolute NMAE differences between the CHARMM36-nbfix~\cite{Huang2017-10.1038/nmeth.4067} results and those obtained from posterior-sampled parametrizations for each species and QoI.
For each molecule, vertical silhouette-style bars illustrate the distribution of per-QoI changes: positive values indicating improvements and negative values indicating regressions relative to the original CHARMM36-nbfix model.

\section{Ca$^{2+}$-troponin binding}

\subsection{cTnC parametrization}

\begin{figure*}
    \centering
    \includegraphics[width=0.75\linewidth]{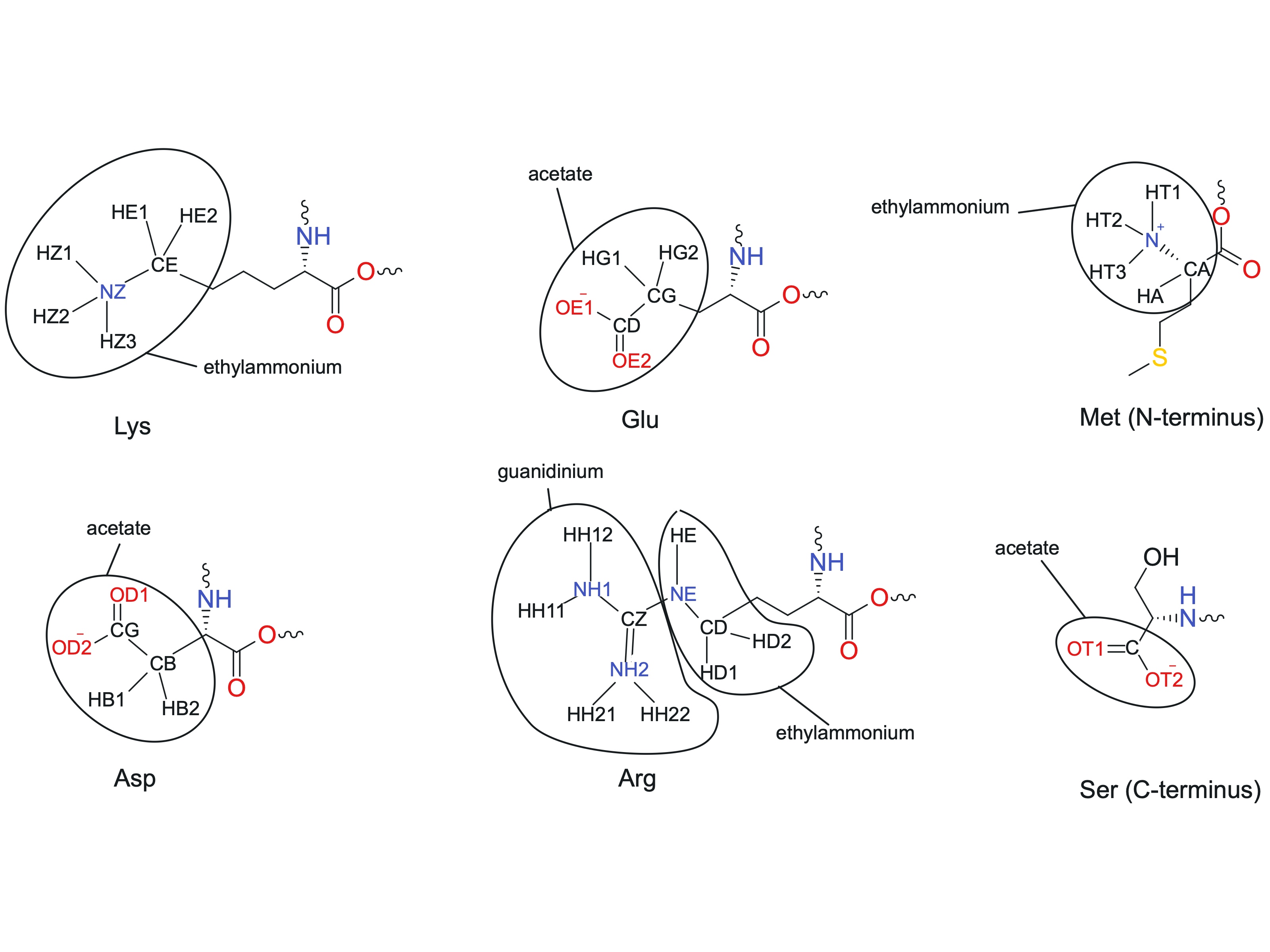}
    \caption{Charged residue of cTnC decomposed into the fragments for parameterization.
    Atoms to-be-parameterized are labeled by the CHARMM36 atom names.}
    \label{fig:aa-params}
\end{figure*}

Partial charges of side chains of amino acids Asp, Glu, Arg and Lys together with the N-terminus (Met) and C-terminus (Ser) were sampled from posterior distributions of their associated fragments.
Decomposition of the residues into the fragments is depicted in Figure~\ref{fig:aa-params}.
Partial charge parameterizations are summarized in Table~\ref{tab:aa-params} for the 10 samples together with the \textit{maximum a posteriori} (MAP).

\begin{figure*}
    \centering
    \includegraphics[width=0.75\linewidth]{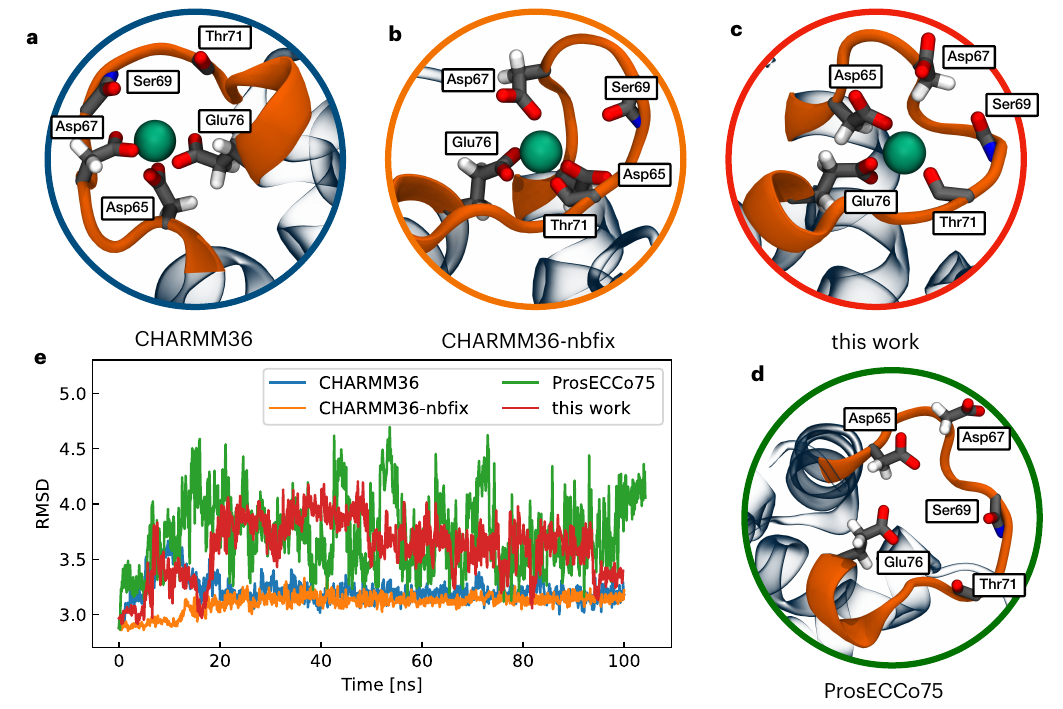}
    \caption{
        \textbf{Equilibrium EF-hand loop of cTnC.}
        \textbf{a-d}: Representative snapshots of the EF-hand loop (orange) of the cTnC (transparent blue), with Ca$^{2+}$-coordinating residues shown in licorice representation and labeled by amino acid code and residue number for each force field.
        \textbf{e}: RMSD of the EF-hand loop as a function of time for the four force fields.
    }
    \label{fig:binding-sites}
\end{figure*}

\begin{table*}
\centering
\caption{Partial charge parameterizations sampled from the Bayesian posterior along with the posterior MAP for the charged cTnC residues.}
\begin{tabular}{ll*{11}{r}}
& & \multicolumn{11}{c}{Posterior sample} \\
& Atom & 1 & 2 & 3 & 4 & 5 & 6 & 7 & 8 & 9 & 10 & MAP \\
\toprule
\multirow{6}{*}{Asp} 
 & OD1 & -0.559 & -0.494 & -0.539 & -0.611 & -0.581 & -0.547 & -0.511 & -0.536 & -0.561 & -0.508 & -0.554 \\
 & OD2 & -0.559 & -0.494 & -0.539 & -0.611 & -0.581 & -0.547 & -0.511 & -0.536 & -0.561 & -0.508 & -0.554 \\
 & CG  &  0.528 &  0.445 &  0.477 &  0.721 &  0.670 &  0.598 &  0.416 &  0.502 &  0.629 &  0.439 & 0.554 \\
 & CB  & -0.340 & -0.373 & -0.241 & -0.495 & -0.328 & -0.384 & -0.158 & -0.340 & -0.365 & -0.353 & -0.330 \\
 & HB1 &  0.065 &  0.058 &  0.021 &  0.098 &  0.010 &  0.040 & -0.018 &  0.055 &  0.029 &  0.065 & 0.042 \\
 & HB2 &  0.065 &  0.058 &  0.021 &  0.098 &  0.010 &  0.040 & -0.018 &  0.055 &  0.029 &  0.065 & 0.042 \\
\midrule
\multirow{6}{*}{Glu} 
 & OE1 & -0.559 & -0.494 & -0.539 & -0.611 & -0.581 & -0.547 & -0.511 & -0.536 & -0.561 & -0.508 & -0.554 \\
 & OE2 & -0.559 & -0.494 & -0.539 & -0.611 & -0.581 & -0.547 & -0.511 & -0.536 & -0.561 & -0.508 & -0.554 \\
 & CD  &  0.528 &  0.445 &  0.477 &  0.721 &  0.670 &  0.598 &  0.416 &  0.502 &  0.629 &  0.439 & 0.554 \\
 & CG  & -0.340 & -0.373 & -0.241 & -0.495 & -0.328 & -0.384 & -0.158 & -0.340 & -0.365 & -0.353 & -0.330 \\
 & HG1 &  0.065 &  0.058 &  0.021 &  0.098 &  0.010 &  0.040 & -0.018 &  0.055 &  0.029 &  0.065 & 0.042 \\
 & HG2 &  0.065 &  0.058 &  0.021 &  0.098 &  0.010 &  0.040 & -0.018 &  0.055 &  0.029 &  0.065 & 0.042 \\
\midrule
\multirow{12}{*}{Arg} 
 & NH1 & -0.761 & -0.671 & -0.792 & -0.771 & -0.731 & -0.738 & -0.679 & -0.675 & -0.757 & -0.842 & -0.765 \\
& HH11 & 0.398 & 0.398 & 0.400 & 0.407 & 0.393 & 0.414 & 0.390 & 0.389 & 0.393 & 0.421 & 0.406 \\
& HH12 & 0.398 & 0.398 & 0.400 & 0.407 & 0.393 & 0.414 & 0.390 & 0.389 & 0.393 & 0.421 & 0.406 \\
& NH2 & -0.761 & -0.671 & -0.792 & -0.771 & -0.731 & -0.738 & -0.679 & -0.675 & -0.757 & -0.842 & -0.765 \\
& HH21 & 0.398 & 0.398 & 0.400 & 0.407 & 0.393 & 0.414 & 0.390 & 0.389 & 0.393 & 0.421 & 0.406 \\
& HH22 & 0.398 & 0.398 & 0.400 & 0.407 & 0.393 & 0.414 & 0.390 & 0.389 & 0.393 & 0.421 & 0.406 \\
& CZ & 0.469 & 0.180 & 0.703 & 0.597 & 0.550 & 0.422 & 0.468 & 0.691 & 0.697 & 0.517 & 0.508 \\
& NE & -0.552 & -0.561 & -0.543 & -0.487 & -0.605 & -0.470 & -0.391 & -0.594 & -0.805 & -0.343 & -0.532 \\
& HE & 0.398 & 0.398 & 0.400 & 0.407 & 0.393 & 0.414 & 0.390 & 0.389 & 0.393 & 0.421 & 0.406 \\
& CD & 0.293 & 0.297 & 0.114 & 0.001 & 0.328 & -0.044 & 0.201 & 0.066 & 0.321 & 0.181 & 0.15 \\
& HD1 & 0.061 & 0.118 & 0.055 & 0.098 & 0.012 & 0.149 & -0.035 & 0.021 & 0.068 & 0.012 & 0.087 \\
& HD2 & 0.061 & 0.118 & 0.055 & 0.098 & 0.012 & 0.149 & -0.035 & 0.021 & 0.068 & 0.012 & 0.087 \\
\midrule
\multirow{7}{*}{Lys} 
& NZ & -0.552 & -0.561 & -0.543 & -0.487 & -0.605 & -0.47 & -0.391 & -0.594 & -0.805 & -0.343 & -0.532 \\
& HZ1 & 0.347 & 0.354 & 0.367 & 0.359 & 0.383 & 0.358 & 0.325 & 0.379 & 0.398 & 0.341 & 0.367 \\
& HZ2 & 0.347 & 0.354 & 0.367 & 0.359 & 0.383 & 0.358 & 0.325 & 0.379 & 0.398 & 0.341 & 0.367 \\
& HZ3 & 0.347 & 0.354 & 0.367 & 0.359 & 0.383 & 0.358 & 0.325 & 0.379 & 0.398 & 0.341 & 0.367 \\
& CE & 0.189 & 0.063 & 0.132 & 0.014 & 0.232 & -0.102 & 0.286 & 0.215 & 0.275 & 0.096 & 0.057 \\
& HE1 & 0.061 & 0.118 & 0.055 & 0.098 & 0.012 & 0.149 & -0.035 & 0.021 & 0.068 & 0.012 & 0.087 \\
& HE2 & 0.061 & 0.118 & 0.055 & 0.098 & 0.012 & 0.149 & -0.035 & 0.021 & 0.068 & 0.012 & 0.087 \\
\midrule
\multirow{6}{*}{Met} 
& N & -0.552 & -0.561 & -0.543 & -0.487 & -0.605 & -0.470 & -0.391 & -0.594 & -0.805 & -0.343 & -0.532 \\
& HT1 & 0.347 & 0.354 & 0.367 & 0.359 & 0.383 & 0.358 & 0.325 & 0.379 & 0.398 & 0.341 & 0.367 \\
& HT2 & 0.347 & 0.354 & 0.367 & 0.359 & 0.383 & 0.358 & 0.325 & 0.379 & 0.398 & 0.341 & 0.367 \\
& HT3 & 0.347 & 0.354 & 0.367 & 0.359 & 0.383 & 0.358 & 0.325 & 0.379 & 0.398 & 0.341 & 0.367 \\
& CA & 0.250 & 0.181 & 0.187 & 0.112 & 0.244 & 0.047 & 0.251 & 0.236 & 0.343 & 0.108 & 0.144 \\
& HA & 0.061 & 0.118 & 0.055 & 0.098 & 0.012 & 0.149 & -0.035 & 0.021 & 0.068 & 0.012 & 0.087 \\
\midrule
\multirow{3}{*}{Ser} 
 & C   &  0.318 &  0.188 &  0.278 &  0.422 &  0.362 &  0.294 &  0.222 &  0.272 &  0.322 &  0.216 & 0.308 \\
 & OT1 & -0.559 & -0.494 & -0.539 & -0.611 & -0.581 & -0.547 & -0.511 & -0.536 & -0.561 & -0.508 & -0.554 \\
 & OT2 & -0.559 & -0.494 & -0.539 & -0.611 & -0.581 & -0.547 & -0.511 & -0.536 & -0.561 & -0.508 & -0.554 \\
\bottomrule
\end{tabular}
\label{tab:aa-params}
\end{table*}

\newpage
\clearpage

\subsection{Binding site}

Starting from the 1AP4 crystal structure~\cite{Spyracopoulos1997-10.1021/bi971223d}, the system was relaxed under four different force fields: CHARMM36/CHARMM36-nbfix~\cite{Huang2017-10.1038/nmeth.4067}, ProsECCo75~\cite{Nencini2024-10.1021/acs.jctc.4c00743}, and the present parameterization—for 100~ns without any restraints.
Figure \ref{fig:binding-sites}e shows the time evolution of the RMSD of the EF-hand loop of cTnC (residues 65–76) relative to the crystal structure.
For each force field, an equilibrated conformation of the EF-hand loop (Figures \ref{fig:binding-sites}a–d) was identified and subsequently restrained for the alchemical double-decoupling free energy calculations.
Note that the RMSD traces do not start from zero because the crystal structure was energy-minimized and pre-equilibrated for 1~ns to relax the solvent environment prior to the production runs.

All force fields except ProsECCo75 retained the Ca$^{2+}$ ion within the binding site over the simulated time, indicating that ProsECCo75 underestimates calcium binding affinity.
Both CHARMM36 and CHARMM36-nbfix produced a stable and rigid binding site, as reflected by the small RMSD fluctuations in Figure \ref{fig:binding-sites}.
In contrast, the present parameterization exhibits larger RMSD fluctuations, suggesting a more flexible binding site.
Because ProsECCo75 does not stabilize the Ca$^{2+}$ ion, it displays the largest RMSD fluctuations among all tested force fields.

The alchemical binding free energy calculations required a well-defined binding site to ensure that the Ca$^{2+}$ ion remained bound during its gradual decoupling.
To this end, selected residues of the EF-hand loop were restrained based on their distances from the Ca$^{2+}$ ion, as determined from the equilibrated snapshots shown in Figure~\ref{fig:binding-sites}a–d.
Carboxylate-containing residues (Asp65, Asp67, and Glu76) were characterized using their carboxylate carbon atoms so as not to bias the restraints toward a specific coordinating oxygen atom.
Residues that coordinate Ca$^{2+}$ through their backbone carbonyl oxygen atoms (Ser69 and Thr71) were characterized directly by those oxygen atoms.

The restrained distances were governed by a flat-bottom harmonic potential defined as
\begin{equation}
    V_\text{fb}(d) = \begin{cases} 0, & \text{if } d_1 \le r \le d_2 \\
    \frac{1}{2} k (d - d_2)^2, & \text{if } r > d_2 \\
    \frac{1}{2} k (d - d_1)^2, & \text{if } r < d_1,
    \end{cases}
\end{equation}
where $d$ is the instantaneous distance, $d_1$ and $d_2$ are the lower and upper bounds of the flat region, respectively, and $k$ is the force constant.
These restraints were implemented using the \texttt{COLVARS}~ module~\cite{Fiorin2013-10.1080/00268976.2013.813594, Fiorin2024-10.1021/acs.jpcb.4c05604} in GROMACS~\cite{Abraham2015-10.1016/j.softx.2015.06.001a}, and the specific parameter values used for each force field are summarized in Table~\ref{tab:fb}.

\begin{table*}[h!]
\centering
\caption{Flat-bottom restraints-based definitions of the binding site under different force fields}
\label{tab:fb}
\begin{tabular}{c c c c c}
\toprule
\multirow{2}{*}{Force field} & Atoms & \multicolumn{3}{c}{flat-bottom restraint} \\
\cmidrule(lr){3-5}
& index (residue) & $d_1$ [nm] & $d_2$ [nm] & $k$ [kJ mol$^{-1}$ nm$^{-2}$] \\
\midrule
\multirow{5}{*}{CHARMM36}  & 1365(Ca$^{2+}$) -- 1004(Asp65) & 0.2 & 0.4 & 200000 \\
                           & 1365(Ca$^{2+}$) -- 1145(Glu76) & 0.2 & 0.4 & 200000 \\
                           & 1365(Ca$^{2+}$) -- 1031(Glu76) & 0.2 & 0.4 & 200000 \\
                           & 1365(Ca$^{2+}$) -- 1053(Ser69) & 0.2 & 0.6 & 200000 \\
                           & 1365(Ca$^{2+}$) -- 1174(Thr71) & 0.2 & 0.6 & 200000 \\
\midrule
\multirow{4}{*}{CHARMM36-nbfix}  & 1365(Ca$^{2+}$) -- 1004(Asp65) & 0.2 & 0.4 & 200000 \\
                                 & 1365(Ca$^{2+}$) -- 1145(Glu76) & 0.2 & 0.4 & 200000 \\
                                 & 1365(Ca$^{2+}$) -- 1031(Glu76) & 0.2 & 0.4 & 200000 \\
                                 & 1365(Ca$^{2+}$) -- 1174(Thr71) & 0.2 & 0.6 & 200000 \\
\midrule
\multirow{5}{*}{ProsECCo75} & 1365(Ca$^{2+}$) -- 1004(Asp65) & 0.2 & 0.4 & 200000 \\
                            & 1365(Ca$^{2+}$) -- 1145(Glu76) & 0.2 & 0.4 & 200000 \\
                            & 1365(Ca$^{2+}$) -- 1031(Glu76) & 0.2 & 0.6 & 200000 \\
                            & 1365(Ca$^{2+}$) -- 1053(Ser69) & 0.2 & 0.6 & 200000 \\
                            & 1365(Ca$^{2+}$) -- 1174(Thr71) & 0.2 & 0.6 & 200000 \\
\midrule
\multirow{4}{*}{this work} & 1365(Ca$^{2+}$) -- 1004(Asp65) & 0.2 & 0.4 & 200000 \\
                           & 1365(Ca$^{2+}$) -- 1145(Glu76) & 0.2 & 0.4 & 200000 \\
                           & 1365(Ca$^{2+}$) -- 1053(Ser69) & 0.2 & 0.6 & 200000 \\
                           & 1365(Ca$^{2+}$) -- 1174(Thr71) & 0.2 & 0.6 & 200000 \\
\bottomrule
\end{tabular}
\end{table*}

\begin{table*}
  \centering
  \caption{
    Overview of reference AIMD trajectories used for each molecule.
    Each system contains a single molecule surrounded by water (and an ion probe).
    Distance between the specified atoms is restrained.
    }
  \label{tab:system_overview_multi}
  \begin{tabular}{@{}l *{7}{c}@{}}
    \toprule
    \multirow{2}{*}{name} & 
    \multirow{2}{*}{composition} & 
    \multirow{2}{*}{length [ps]} & 
    \multicolumn{3}{c}{restraints} &
    \multirow{2}{*}{training samples} &
    \multirow{2}{*}{reference} \\
    \cmidrule(lr){4-6}
    & & & Atoms & $x_0$ [nm] & $k$ [kJ/mol/nm$^2$] & \\
    \midrule
    \multirow{3}{*}{acetaldehyde}
        & 128 waters & 55 & --- & --- & --- & 5000 & this work \\
        & 128 waters + Ca$^{2+}$ & 55 & C(carboxyl)--Ca & 0.35 & 10000 & 5000 & this work \\
        & 128 waters + Ca$^{2+}$ & 55 & C(carboxyl)--Ca & 0.50 & 5000 & 5000 & this work \\
    \midrule
    \multirow{3}{*}{acetate}
        & 128 waters & 55 & --- & --- & --- & 5000 & this work \\
        & 128 waters + Ca$^{2+}$ & 50 & C(carboxyl)--Ca & 0.35 & 20000 & 5000 & this work \\
        & 128 waters + Ca$^{2+}$ & 50 & C(carboxyl)--Ca & 0.50 & 20000 & 5000 & this work \\
    \midrule
    \multirow{3}{*}{acetatamide}
        & 128 waters & 55 & --- & --- & --- & 5000 & this work \\
        & 128 waters + Ca$^{2+}$ & 55 & C(carbonyl)--Ca & 0.35 & 10000 & 5000 & this work \\
        & 128 waters + Ca$^{2+}$ & 55 & C(carbonyl)--Ca & 0.50 & 5000 & 5000 & this work \\
    \midrule
    \multirow{3}{*}{dihydrogenphosphate}
        & 128 waters & 55 & --- & --- & --- & 3935 & this work \\
        & 128 waters + Ca$^{2+}$ & 55 & P--Ca & 0.35 & 5000 & 3935 & this work \\
        & 128 waters + Ca$^{2+}$ & 55 & P--Ca & 0.50 & 5000 & 3935 & this work \\
    \midrule
    \multirow{3}{*}{dimethylether}
        & 128 waters & 30 & --- & --- & --- & 2000 & this work \\
        & 128 waters + Ca$^{2+}$ & 55 & O--Ca & 0.35 & 10000 & 2000 & this work \\
        & 128 waters + Ca$^{2+}$ & 55 & O--Ca & 0.50 & 5000 & 2000 & this work \\
    \midrule
    \multirow{3}{*}{dimethylphosphate}
        & 128 waters & 55 & --- & --- & --- & 5000 & this work \\
        & 128 waters + Ca$^{2+}$ & 50 & P--Ca & 0.35 & 10000 & 5000 & this work \\
        & 128 waters + Ca$^{2+}$ & 50 & P--Ca & 0.50 & 5000 & 5000 & this work \\
    \midrule
    \multirow{3}{*}{ethylammonium}
        & 128 waters & 55 & --- & --- & --- & 10000 & this work \\
        & 128 waters + Cl$^-$ & 50 & N--Cl & 0.35 & 10000 & 10000 & this work \\
        & 128 waters + Cl$^-$ & 50 & N--Cl & 0.50 & 5000 & 10000 & this work \\
    \midrule
    \multirow{3}{*}{ethanol}
        & 128 waters & 55 & --- & --- & --- & 9986 & this work \\
        & 128 waters + Ca$^{2+}$ & 55 & O--Ca & 0.35 & 10000 & 9986 & this work \\
        & 128 waters + Ca$^{2+}$ & 55 & O--Ca & 0.50 & 5000 & 9986 & this work \\
    \midrule
    \multirow{3}{*}{formate}
        & 128 waters & 55 & --- & --- & --- & 2000 & this work \\
        & 128 waters + Ca$^{2+}$ & 60 & O--Ca & 0.325 & 30000 & 2000 & Ref.~\citenum{Martinek2018-10.1063/1.5006779a} \\
        & 128 waters + Ca$^{2+}$ & 60 & O--Ca & 0.525 & 10000 & 2000 & Ref.~\citenum{Martinek2018-10.1063/1.5006779a} \\
    \midrule
    \multirow{3}{*}{guanidinium}
        & 128 waters & 55 & --- & --- & --- & 2000 & this work \\
        & 128 waters + Cl$^-$ & 55 & N--Cl & 0.35 & 5000 & 2000 & this work \\
        & 128 waters + Cl$^-$ & 55 & N--Cl & 0.50 & 5000 & 2000 & this work \\
    \midrule
    \multirow{3}{*}{monohydrogenphosphate}
        & 128 waters & 55 & --- & --- & --- & 4503 & this work \\
        & 128 waters + Ca$^{2+}$ & 55 & P--Ca & 0.35 & 5000 & 4503 & this work \\
        & 128 waters + Ca$^{2+}$ & 55 & P--Ca & 0.50 & 5000 & 4503 & this work \\
    \midrule
    \multirow{3}{*}{methylacetate}
        & 128 waters & 30 & --- & --- & --- & 10000 & this work \\
        & 128 waters + Ca$^{2+}$ & 55 & O(carbonyl)--Ca & 0.35 & 10000 & 10000 & this work \\
        & 128 waters + Ca$^{2+}$ & 55 & O(carbonyl)--Ca & 0.50 & 5000 & 10000 & this work \\
    \midrule
    \multirow{3}{*}{methanol}
        & 128 waters & 45 & --- & --- & --- & 4903 & this work \\
        & 128 waters + Ca$^{2+}$ & 55 & O--Ca & 0.35 & 10000 & 4903 & this work \\
        & 128 waters + Ca$^{2+}$ & 55 & O--Ca & 0.50 & 5000 & 4903 & this work \\
    \midrule
    \multirow{3}{*}{methylsulfate}
        & 128 waters & 55 & --- & --- & --- & 5000 & this work \\
        & 128 waters + Ca$^{2+}$ & 200 & S--Ca & 0.34 & 20000 & 5000 & Ref.~\citenum{Riopedre-Fernandez2024-10.1021/acs.jcim.4c00981} \\
        & 128 waters + Ca$^{2+}$ & 200 & S--Ca & 0.50 & 2500 & 5000 & Ref.~\citenum{Riopedre-Fernandez2024-10.1021/acs.jcim.4c00981} \\
    \midrule
    \multirow{3}{*}{ammonium}
        & 128 waters & 55 & --- & --- & --- & 5000 & this work \\
        & 128 waters + Cl$^-$ & 55 & N--Cl & 0.35 & 5000 & 5000 & this work \\
        & 128 waters + Cl$^-$ & 55 & N--Cl & 0.50 & 5000 & 5000 & this work \\
    \midrule
    \multirow{3}{*}{\textit{N}-methylacetamide}
        & 128 waters & 55 & --- & --- & --- & 10000 & this work \\
        & 128 waters + Ca$^{2+}$ & 55 & O--Ca & 0.35 & 10000 & 10000 & this work \\
        & 128 waters + Ca$^{2+}$ & 55 & O--Ca & 0.50 & 5000 & 10000 & this work \\
    \midrule
    \multirow{3}{*}{\textit{N}-methylsulfamate}
        & 128 waters & 55 & --- & --- & --- & 9521 & this work \\
        & 128 waters + Ca$^{2+}$ & 200 & S--Ca & 0.34 & 10000 & 9521 & Ref.~\citenum{Riopedre-Fernandez2024-10.1021/acs.jcim.4c00981} \\
        & 128 waters + Ca$^{2+}$ & 200 & S--Ca & 0.50 & 2500 & 9521 & Ref.~\citenum{Riopedre-Fernandez2024-10.1021/acs.jcim.4c00981} \\
    \midrule
    \multirow{3}{*}{tetramethylammonium}
        & 128 waters & 55 & --- & --- & --- & 5000 & this work \\
        & 128 waters + Cl$^-$ & 55 & N--Cl & 0.35 & 10000 & 5000 & this work \\
        & 128 waters + Cl$^-$ & 55 & N--Cl & 0.50 & 5000 & 5000 & this work \\
    \bottomrule
  \end{tabular}
\end{table*}

\vspace{1cm}
\section*{References}
%